\documentclass[%
 reprint,
showpacs,
 amsmath,amssymb,
 aps,
 prl
]{revtex4-1}

\usepackage{graphicx}
\usepackage{dcolumn}
\usepackage{bm}

\begin{document}

\preprint{APS/123-QED}

\title{Anomalous Nonlocal Conductance as a Fingerprint of Chiral Majorana Edge States}

\author{Satoshi Ikegaya$^{1}$}
\author{Yasuhiro Asano$^{2,3,4}$}
\author{Dirk Manske$^{1}$}
\affiliation{
$^{1}$Max-Planck-Institut f\"ur Festk\"orperforschung, Heisenbergstrasse 1, D-70569 Stuttgart, Germany\\
$^{2}$Department of Applied Physics, Hokkaido University, Sapporo 060-8628, Japan\\
$^{3}$Center of Topological Science and Technology, Hokkaido University, Sapporo 060-8628, Japan\\
$^{4}$Moscow Institute of Physics and Technology, 141700 Dolgoprudny, Russia}

\date{\today}

\begin{abstract}
Chiral $p$-wave superconductor is the primary example of topological systems hosting chiral Majorana edge states.
Although candidate materials exist, the conclusive signature of chiral Majorana edge states has not yet been observed in experiments.
Here we propose a smoking-gun experiment to detect the chiral Majorana edge states on the basis of theoretical results for
the nonlocal conductance in a device consisting of a chiral $p$-wave superconductor and two ferromagnetic leads.
The chiral nature of Majorana edge states causes an anomalously long-range and chirality-sensitive nonlocal transport in these junctions.
These two drastic features enable us to identify the moving direction of chiral Majorana edge states in the single experimental setup.
\end{abstract}

\maketitle


\textit{Introduction and main idea.}---Superconductors (SCs) with spin-triplet chiral $p$-wave pairing symmetry have attracted intensive attention for the past two decades
because they exhibit topologically protected chiral Majorana edge states (CMESs) having great potential applications to topological quantum computations~\cite{green_00, ivanov_01}.
According to a range of experimental~\cite{maeno_94, maeno_98, ishida_98, maeno_04, maeno_06} and theoretical~\cite{rice_95, yanase_03, wang_13} evidence,
the perovskite superconductor Sr$_2$RuO$_4$ is the most promising candidate for the spin-triplet chiral $p$-wave SCs.
At present, finding a smoking-gun signature of CMESs in this compound is an on-going and central subject in both
physics of topological condensed matter~\cite{kane_r10, zhang_r11, sato_r17} and that of spin-triplet superconductivity~\cite{maeno_r03, maeno_r12, kallin_r12}.

There have been three standard directions for the detection of CMESs.
The first direction is by measurements of internal magnetic fields due to the spontaneous edge current~\cite{sigrist_99, furusaki_01, stone_04, asano_16}.
However, the scanning SQUID experiments for Sr$_2$RuO$_4$ did not detect the expected fields~\cite{moler_05, nelson_07}
because of either the screening currents in the bulk~\cite{sigrist_99} or for other reasons~\cite{sigrist_14, kallin_15, simon_15}.
The second direction is by use of phenomena analogous to the quantum Hall effect in a two-dimensional electron gas with applied magnetic fields~\cite{klitzing_80, thouless_82}:
the spin quantum Hall effect~\cite{fisher_99} and thermal quantum Hall effect~\cite{vishwanath_01}.
However, these effects have not been observed yet because of difficulties in spin and thermal transport measurements.
The third direction studies anomalies in local charge transport of superconducting junctions,
such as a zero-bias conductance peak in tunneling spectroscopy~\cite{kashiwaya_97} and a low-temperature anomaly in Josephson currents~\cite{asano_02}.
However, roughly speaking, these anomalies can be induced by any type of mid-gap Andreev bound states and are not unique to the CMESs.
Therefore, unfortunately, the zero-bias conductance peak observed in a planar tunneling experiment for Sr$_2$RuO$_4$~\cite{kashiwaya_11}
cannot be the conclusive evidence for the CMESs.

\begin{figure}[tttt]
\begin{center}
\includegraphics[width=0.32\textwidth]{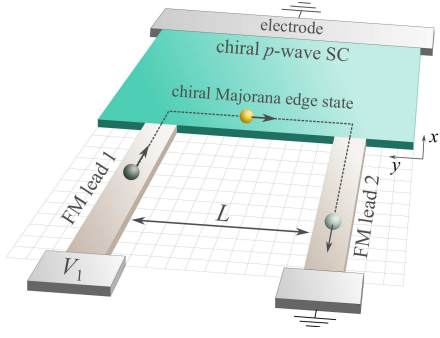}
\caption{Schematic image of the device consisting of a chiral $p$-wave superconductor with two ferromagnetic leads.
The figure corresponds to the situation for measuring $G_{21}$, where the bias voltage $V_{1}$ is applied to the electrode attached the lead $1$,
while the electrodes attached to the lead $2$ and superconductor are grounded.
}
\label{fig:figure1}
\end{center}
\end{figure}

To resolve this stalemate, in the present Letter, we propose a novel experiment that provide a smoking-gun signature of CMESs though charge transport measurements.
The central ingredient of our scheme is that we measure \textit{nonlocal} charge transport in the presence of CMESs~\cite{beenakker_10}.
We will use a setup as shown in Fig.~\ref{fig:figure1}, where two ferromagnetic (FM) leads are attached to an edge of a chiral $p$-wave SC~\cite{tserkovnyak_18}.
The nonlocal conductance in a similar device with replacing the chiral $p$-wave SC by a conventional $s$-wave SC has been already studied~\cite{deutscher_00, yamashita_03}.
In such a device the nonlocal conductance is governed by two distinctive nonlocal transport processes yielding opposite contributions:
an incident electron from one lead is scattered into another lead as an electron (elastic co-tunneling process) or a hole (crossed Andreev reflection process).
The exchange potential in the FM leads is source of finite nonlocal conductance because it generates the imbalance between these two nonlocal transport processes
~\cite{deutscher_00, yamashita_03}.
With conventional $s$-wave SC, the subgap nonlocal conductance is strongly suppressed when the distance between the two leads exceeds the superconducting coherent length.
This is because that incident electrons must tunnel from one lead to the other through evanescent waves of Bogoliubov quasi-particles in the superconducting segment.
However, we expect that CMESs modify the situation drastically such that the CMESs moving in the direction from lead $\alpha$ to $\beta$
mediate the nonlocal transport from lead $\alpha$ to $\beta$ irrespective of the distance between the two leads,
while it does not assist the nonlocal transport from lead $\beta$ to $\alpha$ (See also Fig.~\ref{fig:figure1}).
If we can capture such unusual anisotropy in the nonlocal transport processes, it can be a smoking-gun signature of the CMESs.

\begin{figure}[tttt]
\begin{center}
\includegraphics[width=0.5\textwidth]{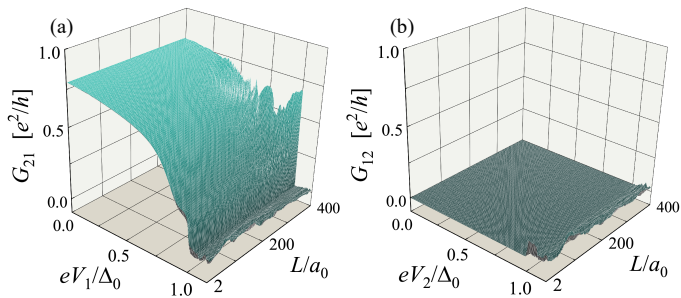}
\caption{ Nonlocal conductance (a) $G_{2 1}$ and (b) $G_{1 2}$ are plotted as
a function of the bias voltage and distance between the FM leads $L$.
We vary $L$ from $0.2\xi_0$ to $40\xi_0$.
The spectrum of $G_{21}$ and that of $G_{12}$ are related to each other by changing the moving direction of the chiral Majorana edge states.}
\label{fig:figure2}
\end{center}
\end{figure}

We calculate two types of nonlocal differential conductance $G_{21} = dI_2/dV_1$ and $G_{12} = dI_1/dV_2$ by using the lattice Green function technique.
Here, $I_{\alpha}$ is the current response in the FM lead $\alpha$ due to the application of the bias voltage $V_\beta$ to the electrode attached to the FM lead $\beta$,
while the electrodes attached to the FM lead $\alpha$ and superconductor are grounded.
We will demonstrate that spectrum of $G_{21}$ and $G_{12}$ indeed exhibit the distinctive contrast reflecting the chiral motion of CMESs.
Namely, when the CMESs move in the direction from the lead $\alpha$ to $\beta$,
the nonlocal conductance $G_{\beta \alpha}$ becomes finite irrespective of the distance between the FM leads [See Fig.~\ref{fig:figure2} (a)], 
while the nonlocal conductance $G_{\alpha \beta}$ becomes almost zero [See Fig.~\ref{fig:figure2} (b)].
We can measure both  $G_{21}$ and $G_{12}$ only by changing the lead wire to which the bias-voltage is applied.
Therefore, we can identify the moving direction of the CMES in the single experimental setup.
The remarkable advantage of our proposal is that
we only need the obvious difference in $G_{21}$ and $G_{12}$, where one of them is \textit{finite} and the other is \textit{zero},
to identify the CMESs in the chiral $p$-wave superconductor conclusively.

%
\textit{Minimal model.}---Let us consider the junction illustrated in Fig.~\ref{fig:figure1} on a two dimensional tight-binding model with the lattice constant $a_0$.
A lattice site is indicated by a vector $\boldsymbol{r}=j\boldsymbol{x}+m\boldsymbol{y}$, 
where $\boldsymbol{x}$ ($\boldsymbol{y}$) is the vector in the $x$ ($y$) direction with $|\boldsymbol{x}|= |\boldsymbol{y}|=a_0$.
The chiral $p$-wave SC occupies $j \geq1$ and $- M_\mathrm{s} \leq m \leq M_\mathrm{s}$, where its width is given by $W_\mathrm{s}/a_0 =  2 M_\mathrm{s}$.
In the $y$ direction, we apply the hard-wall boundary condition.
The FM lead 1 (FM lead 2) is placed on $j \leq 0$ and $m_\mathrm{f} \leq m \leq M_\mathrm{f}$ ($-m_\mathrm{f} \geq m \geq -M_\mathrm{f}$),
where its width is denoted by $W_\mathrm{f}/a_0 = M_\mathrm{f} - m_\mathrm{f}$.
The distance between the two FM leads is given by $L/a_0 = 2 m_\mathrm{f}$.
The present device is described by the Bogoliubov-de Gennes Hamiltonian
$H = H_\mathrm{s} + H_\mathrm{1} + H_\mathrm{2}$.
In this paper, we phenomenologically describe the chiral $p$-wave SC by using the standard minimal model
\begin{align}
H_\mathrm{s} = \frac{1}{2}\sum_{\boldsymbol{r}, \boldsymbol{r}^{\prime}}
\boldsymbol{c}^{\dagger}_{\boldsymbol{r}}
\left[ \begin{array}{cc}
\hat{\xi}^\mathrm{s}_{\boldsymbol{r}, \boldsymbol{r}^{\prime}} & \hat{\Delta}_{\boldsymbol{r}, \boldsymbol{r}^{\prime}} \\
-\hat{\Delta}^{\ast}_{\boldsymbol{r}, \boldsymbol{r}^{\prime}} & -\hat{\xi}^\mathrm{s}_{\boldsymbol{r}, \boldsymbol{r}^{\prime}} \\
\end{array}\right]
\boldsymbol{c}_{\boldsymbol{r}^{\prime}},
\end{align}
where $j, j^{\prime} > 0$,
$\hat{\xi}^\mathrm{s}_{\boldsymbol{r}, \boldsymbol{r}^{\prime}} =
[-t_\mathrm{s} \delta_{|\boldsymbol{r}-\boldsymbol{r}^{\prime}|, a_0} + (4t_\mathrm{s}-\mu_\mathrm{s}) \delta_{\boldsymbol{r},\boldsymbol{r}^{\prime}}] \hat{\sigma}_0$,
$\hat{\Delta}_{\boldsymbol{r}, \boldsymbol{r}^{\prime}} =
(\Delta_0/2) [ i(\delta_{j,j^{\prime}+1} - \delta_{j+1,j^{\prime}}) - \chi (\delta_{m,m^{\prime}+1} - \delta_{m+1,m^{\prime}})] \hat{\sigma}_x$,
and $\boldsymbol{c}_{\boldsymbol{r}}
= [c_{\boldsymbol{r},\uparrow},c_{\boldsymbol{r},\downarrow}, c^{\dagger}_{\boldsymbol{r},\uparrow}, c^{\dagger}_{\boldsymbol{r},\downarrow}]^{\mathrm{T}}$
with $c_{\boldsymbol{r},\sigma}^{\dagger}$($c_{\boldsymbol{r},\sigma}$) representing
the creation (annihilation) operator of an electron at the site $\boldsymbol{r}$ with spin $\sigma$ ($ = \uparrow$ or $\downarrow )$.
The Pauli matrices in spin space are represented by $\hat{\sigma}_{i}$ for $i = x, y, z$, and the $2 \times 2$ unit matrix is denoted with $\hat{\sigma}_{0}$.
 $t_\mathrm{s}$ and $\mu_\mathrm{s}$ respectively denote the nearest-neighbor hopping integral and chemical potential in the superconductor.
The amplitude and chirality of the pair potential are represented by $\Delta_0$ and $\chi$ ($=1$ or $-1$), respectively.
The pair potential for a spin-triplet pairing symmetry in momentum space is
generally described as $\hat{\Delta} (\boldsymbol{k}) = \boldsymbol{d}(\boldsymbol{k}) \cdot \hat{\boldsymbol{\sigma}} (i \sigma_y)$.
In this Letter, we use the $d$-vector of $\boldsymbol{d}(\boldsymbol{k}) = \Delta_0 \hat{z} [\sin ( k_x a_0) + i \chi \sin( k_y a_0)] $,
which is the most probable one in Sr$_2$RuO$_4$~\cite{rice_95, yanase_03, maeno_r03, maeno_r12, kallin_r12}.
Here, $k_x$ ($k_y$) represents the wave number along the $x$ ($y$) direction, and $\hat{z}$ represents the unit vector in the $z$-direction corresponding to the $c$-axis of Sr$_2$RuO$_4$.
The FM lead $\alpha$ ($=1$, $2$) is described by
\begin{align}
 H_{\alpha} &=\frac{1}{2}\sum_{\boldsymbol{r}, \boldsymbol{r}^{\prime}}
\boldsymbol{c}^{\dagger}_{\boldsymbol{r}}
\left[ \begin{array}{cc}
\hat{\xi}^\mathrm{\alpha}_{\boldsymbol{r}, \boldsymbol{r}^{\prime}} & 0 \\
0 & -\hat{\xi}^\mathrm{\alpha}_{\boldsymbol{r}, \boldsymbol{r}^{\prime}} \\
\end{array}\right]
\boldsymbol{c}_{\boldsymbol{r}^{\prime}},
\end{align}
where $j \leq 0$, $\hat{\xi}^\mathrm{\alpha}_{\boldsymbol{r}, \boldsymbol{r}^{\prime}} =
[-t_\mathrm{f} \delta_{|\boldsymbol{r}-\boldsymbol{r}^{\prime}|, a_0} + (4t_\mathrm{f}-\mu_\mathrm{f}) \delta_{\boldsymbol{r},\boldsymbol{r}^{\prime}}] \hat{\sigma}_0
+ \boldsymbol{M}_{\alpha}\cdot \hat{\boldsymbol{\sigma}} \delta_{\boldsymbol{r},\boldsymbol{r}^{\prime}} $.
The nearest-neighbor hopping integral and chemical potential in the FM leads are respectively denoted $t_\mathrm{f}$ and $\mu_\mathrm{f}$. 
The exchange potential in the FM lead $\alpha$ is given by
$\boldsymbol{M}_{\alpha} = M_{\alpha} (\cos \theta_{\alpha}  \sin \varphi_{\alpha}, \sin \theta_{\alpha} \sin \varphi_{\alpha}, \cos \varphi_{\alpha})$.
In what follows, we fix several parameters as
$\mu_\mathrm{f}=1.0t_\mathrm{f}$, $t_\mathrm{s}=1.0t_\mathrm{f}$, $\mu_\mathrm{s}=2.0t_\mathrm{f}$, $\Delta = 0.1t_\mathrm{f}$, and $\chi = -1$.
In the tight-binding model, the superconducting coherent length is given by $\xi_0 = (t_\mathrm{s}/\Delta_0)a_0$~\cite{lee_10}.
With our parameter choice, we obtain $\xi_0 = 10 a_0$.
The chiral $p$-wave SC hosts two CMESs originated from the two different spin-sectors.
With $\chi = -1$, both of them move along the edge at $j=1$ in the direction from the FM lead 1 to 2.

We are interested in the nonlocal differential conductance $G_{2 1} (eV_1) = dI_2/dV_1$ and $G_{1 2} (eV_2) = dI_1/dV_2$.
On the basis of the Blonder-Tinkham-Klapwijk (BTK) formalism~\cite{klapwijk_82}, the nonlocal conductance at zero temperature is given by
~\cite{deutscher_00,yamashita_03,pachos_08,pablo_15,mukerjee_17,alidoust_17,akhmerov_18,cheng_18}
\begin{gather}
G_{\beta \alpha} (eV_\alpha) = \frac{e^{2}}{h} 
\left[ - R^{\mathrm{EC}}_{\beta \alpha} + R^{\mathrm{CAR}}_{\beta \alpha} \right]_{eV_\alpha=E}\label{eq:nlg},\\
R^{\mathrm{EC}(\mathrm{CAR})}_{\beta \alpha} = \sum_{\zeta,\eta} | r^{ee (\mathrm{he})}_{\beta \alpha}(\zeta;\eta) |^{2},
\end{gather}
with $\alpha \neq \beta$.
The elastic co-tunneling (EC) and crossed Andreev reflection (CAR) coefficients at energy $E$ are respectively denoted by
$r^{ee}_{\alpha \beta}(\zeta;\eta)$ and $r^{he}_{\alpha \beta}(\zeta;\eta)$,
where the index $\zeta$ ($\eta$) labels the outgoing (incoming) channel in the FM lead $\beta$ (FM lead $\alpha$).
These reflection coefficients are obtained by using the lattice Green function technique~\cite{fisher_81,ando_91}
(See Supplemental Material for the detailed calculation).
In the BTK formalism, we assume that all currents following towards $x = +\infty$ ($x = -\infty$) in the superconductor (FM lead $\beta$)
are absorbed into the ideal electrode which is not describe in the Hamiltonian explicitly.
We note that the BTK formalism is quantitatively justified for bias voltages well below the superconducting gap.
%

%
\textit{Results on nonlocal conductance.}---We first focus on the nonlocal conductance $G_{21}$.
In Fig.~\ref{fig:figure2}(a), we show $G_{21}$ as a function of the bias voltage and distance between the FM leads $L$.
We choose the parameters as $W_\mathrm{f} = 20a_0$ and $W_\mathrm{s} = 500a_0$.
We vary $L$ from $0.2\xi_0$ to $40\xi_0$, where $\xi_0 = 10 a_0$.
For the FM leads, we consider the antiparallel magnetization along $z$ axis, where
$\boldsymbol{M}_{1 (2)} = + (-) M_{ex} \hat{z}$ with $M_{ex}= 0.5 t_\mathrm{f}$.
We find that $G_{21}$ for $eV \ll \Delta_0$ is almost independent of $L$ and is finite for $L \gg \xi_0$.
Specifically, at zero-bias voltage, we find $G_{21} \approx 0.79 (e^2 / h)$ irrespective of $L$.
The anomalously long-range nonlocal transport in the present junction suggests
that wave functions in the two different FM leads are mediated not by evanescent waves but by the propagating waves of CMESs.
We will later confirm this statement by analyzing the wave functions in the present junction.
Next, we discuss the nonlocal conductance $G_{12}$.
In Fig.~\ref{fig:figure2}(b), we show $G_{12}$ as a function of the bias voltage and $L$,
where the parameters are chosen as same as those in Fig.~\ref{fig:figure2}(a).
In contrast to $G_{21}$, we find that $G_{12}$ with $eV < \Delta_0$ is almost zero for all $L$.
This suggests that the CMESs moving in the direction from the lead 1 to 2 cannot assist the nonlocal transport processes from the lead 2 to 1.
In the BTK formalism, we assume that the CMESs moving towards $x=+\infty$ is absorbed into the ideal electrode attached to the superconductor.
To support this assumption, we also calculate the reflection and transmission probabilities at an ideal chiral $p$-wave SC/normal-metal interface,
and confirm that the incident CMESs are always scattered into the attached normal-metal(See Supplemental Material for the detailed calculation).
We confirm that the spectrum of $G_{21}$ and that of $G_{12}$ are replaced each other by changing the sign of chirality from $-1$ to $+1$.
Thus, the distinctive contrast between $G_{21}$ and $G_{12}$ is indeed related with the moving direction of the CMESs.
We can measure both $G_{21}$ and $G_{12}$ by changing the FM lead wire to which the bias voltage is applied.
Therefore, by comparing $G_{21}$ and $G_{12}$,
we can test the sign of chirality, and therefore the moving direction of CMESs, in the single experimental setup.
%

\begin{figure}[bbbb]
\begin{center}
\includegraphics[width=0.5\textwidth]{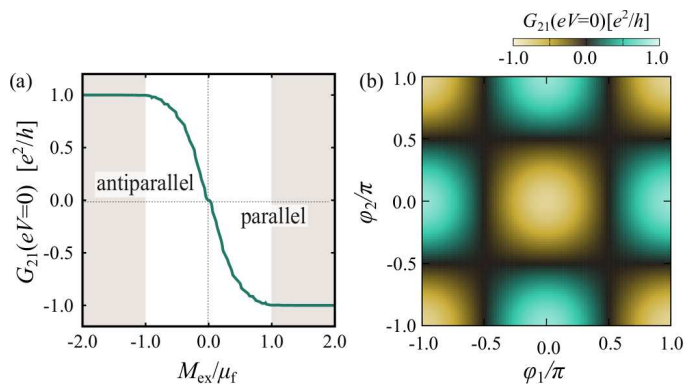}
\caption{(a) Nonlocal conductance $G_{21}$ at zero-bias voltage as a function of the exchange potential $M_{ex}$.
(b) $G_{21}$ at zero-bias voltage as a function of the angles of magnetic moments $\varphi_1$ and $\varphi_2$.}
\label{fig:figure3}
\end{center}
\end{figure}

We now discuss the exchange potential dependence of the nonlocal conductance.
In Fig.~\ref{fig:figure3}(a), we show the nonlocal conductance $G_{21}$ at zero-bias voltage as a function of the exchange potential amplitude.
We here consider either parallel or antiparallel alignment of magnetization along the $z$-axis with $\boldsymbol{M}_{1} = |M_{ex}| \hat{z}$ and $\boldsymbol{M}_{2} = M_{ex}\hat{z}$.
With this representation, the parallel (antiparallel) alignments of the magnetization is described with $M_{ex}>0$ ($M_{ex}<0$).
We choose the parameters as $W_\mathrm{f} = 20a_0$, $W_\mathrm{s} = 500a_0$ and $L=300a_0$.
For the antiparallel ($M_{ex}<0$) and parallel ($M_{ex}>0$) magnetization, $G_{21}$ respectively becomes positive and negative finite, which leads to the relation of
$R^{\mathrm{EC}}_{21} < R^{\mathrm{CAR}}_{21}$ ($R^{\mathrm{EC}}_{21} > R^{\mathrm{CAR}}_{21}$) with the antiparallel (parallel) magnetization.
When the $d$-vector in the superconductor is parallel or antiparallel to the magnetic moment in the FM leads,
Andreev reflection occurs between electron and hole states with opposite spins, while normal reflection occurs between equal-spin electrons~\cite{yoshida_99, hirai_01, hirai_03}.
Therefore, the antiparallel magnetization in the FM leads suppresses the equal-spin scattering process of EC, while it does not damage the CAR process.
On the other hand, the parallel magnetization in the FM leads does not damage the EC process, while it disturbs the spin flip in the CAR process.
This roughly explains the relation of $R^{\mathrm{EC}}_{21} < R^{\mathrm{CAR}}_{21}$ ($R^{\mathrm{EC}}_{21} > R^{\mathrm{CAR}}_{21}$) with the antiparallel (parallel) magnetization.
In the absence of the exchange potential ($M_{ex}=0$), the nonlocal conductance $G_{21}$ becomes zero
due to the complete cancellation between the contributions from the EC and CAR processes (i.e., $R^{\mathrm{EC}}_{21} = R^{\mathrm{CAR}}_{21}$).
When $|M_{ex}|$ exceeds $\mu_{\mathrm{f}}$, only the spin-$\downarrow$ states remain at  the Fermi level  in the FM lead 1
and the only spin-$\uparrow$ (-$\downarrow$) states remain at the Fermi level  in the FM lead 2 with the antiparallel (parallel) alignment of magnetization.
Within such half-metallic limit ($|M_{ex}|>\mu_{\mathrm{f}}$), 
we obtain $G_{21} \approx + (-) e^2/h$ with the antiparallel (parallel) magnetization.
In Fig.~\ref{fig:figure3}(b), we show $G_{21}$ at zero-bias voltage for various directions of the magnetization.
The exchange potentials in the FM lead 1 and FM lead 2 are respectively chosen as
$\boldsymbol{M}_{1} = M_{ex} (\sin \varphi_{1}, 0, \cos \varphi_{1})$ and
$\boldsymbol{M}_{2} = M_{ex} (0, \sin \varphi_{2}, \cos \varphi_{2})$ with $M_{ex}= 0.5 t_\mathrm{f}$.
By changing $\varphi_{1}$ and $\varphi_{2}$, $\boldsymbol{M}_{1}$ and $\boldsymbol{M}_{2}$ are respectively  rotated around the $y$ and $x$ axis.
We choose the parameters as $W_\mathrm{f} = 20a_0$, $W_\mathrm{s} = 300a_0$ and $L=160a_0$.
Except for $\varphi_{1} = \pm \pi/2$ and $\varphi_{2} = \pm \pi/2$, we obtain the finite nonlocal conductance $G_{21}$.
The sign of $G_{21}$ is determined by $- \mathrm{sgn} ({M}_{1}^z) \mathrm{sgn} ({M}_{2}^z)$, where ${M}_{\alpha}^z = M_{ex} \cos \varphi_{\alpha}$.
The maximum magnitude of $G_{21}$ is obtained when both $\boldsymbol{M}_{1}$ and $\boldsymbol{M}_{2}$ are directed along either $+\hat{z}$ or $-\hat{z}$.
We also confirm that the nonlocal conductance $G_{12}$ is zero irrespective of $\varphi_{1}$ and $\varphi_{2}$ for $L \gg \xi_0$.
Therefore, we can find the distinctive contrast in $G_{21}$ and $G_{12}$, which is the evidence for the CMESs, for the various alignments of the magnetization.

\begin{figure}[tttt]
\begin{center}
\includegraphics[width=0.5\textwidth]{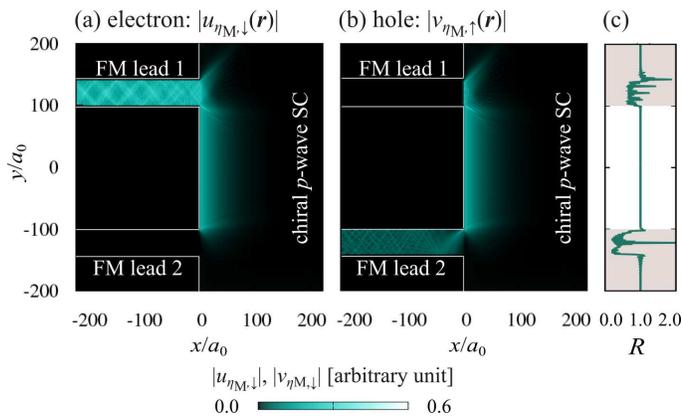}
\caption{Spatial profile of the wave function having the largest contribution to $R^{\mathrm{CAR}}_{21}$.
In (a) and (b), we respectively show the amplitude of electron component $|u_{\eta_{\mathrm{M}},\downarrow}(\boldsymbol{r})|$
and that of hole component $|v_{\eta_{\mathrm{M}},\uparrow}(\boldsymbol{r})|$.
In (c), the ratio of $R=|u_{\eta_{\mathrm{M}},\downarrow}| / |v_{\eta_{\mathrm{M}},\uparrow}|$ at the edge of the superconductor ($j=1$) is plotted as a function of $y$.}
\label{fig:figure4}
\end{center}
\end{figure}

\textit{Majorana wave functions.}---The anomalously long-range nonlocal transport in the present junction implies
that an incident electron from one lead is transmitted through the superconducting segment as the CMESs, and is scattered into another leads.
To confirm this statement directly, we here analyze the quasi-particle wave functions contributing to the CAR process from the FM lead 1 to 2.
Specifically, we calculate the wave function $\psi_{\eta_{\mathrm{M}}}(\boldsymbol{r})
= [u_{\eta_{\mathrm{M}},\uparrow}(\boldsymbol{r}), u_{\eta_{\mathrm{M}},\downarrow}(\boldsymbol{r}),
v_{\eta_{\mathrm{M}},\uparrow}(\boldsymbol{r}), v_{\eta_{\mathrm{M}},\downarrow}(\boldsymbol{r})]^{\mathrm{T}}$ at zero energy,
where $\eta_{\mathrm{M}}$ labels the incoming channel having  the largest contribution to $R^{\mathrm{CAR}}_{21}$
(i.e., $\eta_{\mathrm{M}}$  have the largest value of $\sum_{\zeta} | r^{he}_{21}(\zeta;\eta) |^{2}$ among all $\eta$).
Details for the calculation are given in Supplemental Material.
To discuss the most comprehensible case, we assume the half-metallic ferromagnets with the anitiprallel magnetization along $z$-axis,
where $\boldsymbol{M}_{1 (2)} = + (-) M_{ex} \hat{z}$ with $M_{ex}= 1.5 \mu_\mathrm{f}$.
With this specific choice of magnetization,
$\psi_{\eta_{\mathrm{M}}}(\boldsymbol{r})$ consists of only 
spin-$\downarrow$ electron component $u_{\eta_{\mathrm{M}},\downarrow}$
and spin-$\uparrow$ hole component $v_{\eta_{\mathrm{M}},\uparrow}$,
while $u_{\eta_{\mathrm{M}},\uparrow}=v_{\eta_{\mathrm{M}},\downarrow}=0$.
Moreover, the local Andreev reflection in the FM lead 1 and EC from the FM lead 1 to 2 are absent.
In Fig.~\ref{fig:figure4}(a) and (b), we respectively show the spatial profile of electron component amplitude 
$|u_{\eta_{\mathrm{M}},\downarrow}|$ and that of hole component amplitude $|v_{\eta_{\mathrm{M}},\uparrow}|$.
We choose the parameters as $W_{\mathrm{f}} = 30 a_0$, $W_{\mathrm{s}} = 400 a_0$, and $L=200 a_0$. 
In the lead 1, we find the finite $|u_{\eta_{\mathrm{M}},\downarrow}|$ which corresponds to the incident electron wave and normal-reflected electron wave.
In the lead 2, we find the finite $|v_{\eta_{\mathrm{M}},\uparrow}|$ corresponding to the crossed Andreev reflected hole wave.
There are no propagating hole (electron) waves in the lead 1 ( lead 2) due to the absence of the local Andreev reflection (EC) process.
For the superconducting segment, most importantly, we find that the wave function localized at the edge of the superconductor mediates the wave functions in the two different FM leads.
To examine this in more detail, in Fig.~\ref{fig:figure4}(c), we show the ratio of
$R=|u_{\eta_{\mathrm{M}},\downarrow}| / |v_{\eta_{\mathrm{M}},\uparrow}|$ at the edge of the superconductor ($j=1$).
We find that $R=1.0$ holds between the two FM leads ($-100<m<100$).
Therefore, the wave function bridging the two FM leads indeed corresponds to a Majorana edge excitation
described by the superposition of an electron wave and a hole wave with equal amplitude.

\textit{Discussion}---Here we highlight the most significant advantage of our proposal that
we can identify the CMESs thorough the obvious difference in $G_{21}$ and $G_{12}$, where one of them is \textit{finite} and the other is \textit{zero}.
In real experiments, several perturbations such as the tilt of $d$-vector, and spin-orbit coupling potentials in the vicinity of junction interface,
may induce additional spin-flip scattering processes and may decrease the amplitude of the finite nonlocal conductance.
Even so, our proposal is still valid in the presence of such perturbations because we only need the contrast between \textit{finite} and \textit{zero} nonlocal conductance for detecting the CMESs.
Actually, we have confirmed that the significant contrast between $G_{21}$ and $G_{12}$ is preserved for the broad range of magnetization alignments as shown in Fig.~\ref{fig:figure3}(b).

In summary, we have discussed the nonlocal conductance in the junction consisting of a chiral $p$-wave SC and two FM leads.
The CMESs cause the the anomalously long-range and chirality-sensitive nonlocal transport and generate the drastic contrast in $G_{21}$ and $G_{12}$.
On the basis of these numerical results, we have proposed a smoking-gun experiment to detect the CMESs in chiral $p$-wave superconductors and have discussed the advantage of our proposal.
We hope that our work will motivate further experiments on nonlocal transport measurements for recently fabricated ferromagnetic-SrRuO$_3$/Sr$_2$RuO$_4$ hybrid systems~\cite{maeno_16}.

\begin{acknowledgments}
We are grateful to J. Annett and D. Schlom for fruitful discussions.
YA is supported by ``Topological Materials Science'' (Nos. JP15H05852 and JP15K21717) from
the Ministry of Education, Culture, Sports, Science and Technology (MEXT) of Japan, JSPS Core-to-Core Program(A. Advanced Research Networks), 
Japanese-Russian JSPS-RFBR project (Nos. 2717G8334b and 17-52-50080),
and the Ministry of Education and Science of the Russian Federation (Grant No. 14Y.26.31.0007).
\end{acknowledgments}

\pagebreak
\onecolumngrid
\begin{center}
  \textbf{\large Supplemental Material for \\ ``Anomalous Nonlocal Conductance as a Fingerprint of Chiral Majorana Edge States''}\\ \vspace{0.3cm}
Satoshi Ikegaya$^{1}$, Yasuhiro Asano$^{2,3,4}$, and Dirk Manske$^{1}$\\ \vspace{0.1cm}
{\itshape $^{1}$Max-Planck-Institut f\"ur Festk\"orperforschung, Heisenbergstrasse 1, D-70569 Stuttgart, Germany\\
$^{2}$Department of Applied Physics, Hokkaido University, Sapporo 060-8628, Japan\\
$^{3}$Center of Topological Science and Technology, Hokkaido University, Sapporo 060-8628, Japan\\
$^{4}$Moscow Institute of Physics and Technology, 141700 Dolgoprudny, Russia}
\date{\today}
\end{center}

\section{Recursive Green Function Technique}

\subsection{Reflections Coefficients}

\begin{figure}[hhhh]
\begin{center}
\includegraphics[width=0.6\textwidth]{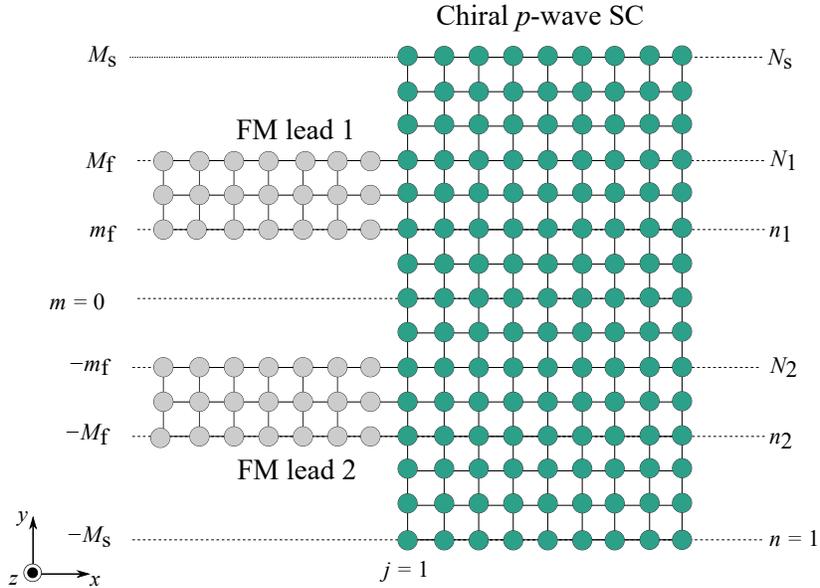}
\caption{Device consisting of a chiral $p$-wave superconductor and two ferromagnetic leads as described within the tight-binding model.
In the main text, a lattice site is indicated by $\boldsymbol{r}=j\boldsymbol{x}+m\boldsymbol{y}$, where $-M_{\mathrm{s}} \leq m \leq M_{\mathrm{s}}$
(See left-side of the figure).
In this supplemental material, for later convenience, we alternatively use $\boldsymbol{r}=j\boldsymbol{x}+n\boldsymbol{y}$ with $1 \leq n \leq N_{\mathrm{s}}$
(See right-side of the figure).
}
\label{fig:supplement}
\end{center}
\end{figure}

In this section we explain the lattice Green function technique used for calculating the reflection coefficients in Eq.~(4) in the main text.
For later convenience, we introduce several integer numbers as (see also Fig.~\ref{fig:supplement})
\begin{gather}
N_\mathrm{s} = 2 M_\mathrm{s} + 1, \quad
N_\mathrm{f} = M_\mathrm{f}-m_\mathrm{f}+ 1, \quad
N_L = 2 m_\mathrm{f} - 1, \quad
N_{L^{\prime}} = M_\mathrm{s} - M_\mathrm{f}, \nonumber\\
n_2 = N_{L^{\prime}}+1, \quad
N_2 = n_2 + N_\mathrm{f}-1, \quad
n_1 = N_2+N_L+1, \quad
N_1 = n_1 + N_\mathrm{f}-1, \nonumber\\
A = 2 N_L, \quad
B = 2 N_{L^{\prime}}, \nonumber
\end{gather}
and indicate a lattice site by a vector $\boldsymbol{r}=j\boldsymbol{x}+n\boldsymbol{y}$ .
The chiral $p$-wave superconductor (SC) occupies $j > 0$ and $1 \leq n \leq N_\mathrm{s}$
and the ferromagnetic (FM) lead $\alpha$ ($=1$-$2$) occupies $j \leq 0$ and $n_{\alpha} \leq n \leq N_{\alpha}$.
We rewrite the Bogoliubov-de Gennes (BdG) Hamiltonian in the main text into the appropriate form for the numerical calculation as
\begin{gather}
H = \sum_{j = -\infty}^{\infty}
\left[ \boldsymbol{C}_{j+1}^{\dagger} \check{T}(j) \boldsymbol{C}_{j}
+ \boldsymbol{C}_{j}^{\dagger} \check{T}^{\dagger}(j) \boldsymbol{C}_{j+1}
+  \boldsymbol{C}_{j}^{\dagger} \check{H} (j) \boldsymbol{C}_{j}\right], \\
\boldsymbol{C}_{j} = [ \boldsymbol{c}_{j}, \boldsymbol{c}^{\dagger}_{j} ]^{\mathrm{T}}, \quad
\boldsymbol{c}_{j} = [ \boldsymbol{c}_{j,1}, \boldsymbol{c}_{j,2}, \cdots, \boldsymbol{c}_{j,N_\mathrm{s}}]^{\mathrm{T}},\quad
\boldsymbol{c}_{j,n} = [ c_{j,n,\uparrow}, c_{j,n,\downarrow}]^{\mathrm{T}},\\
\check{T}(j)=\left\{ \begin{array}{cl} 
\check{T}_{\mathrm{f}} & \text{for}\quad j \leq 0 \\
\check{T}_{\mathrm{s}} & \text{for}\quad j > 0
\end{array}\right., \quad
\check{H}(j)=\left\{ \begin{array}{cl} 
\check{H}_{\mathrm{f}} & \text{for}\quad j \leq 0 \\
\check{H}_{\mathrm{s}} & \text{for}\quad j > 0,
\end{array}\right.,
\end{gather}
where
\begin{gather}
\check{T}_{\mathrm{f}} = \left[ \begin{array}{cc}
\hat{T}_{\mathrm{f}} & 0 \\
0 & - \hat{T}_{\mathrm{f}} \\
\end{array}\right], \quad
\hat{T}_{\mathrm{f}} = \left[ \begin{array}{ccccc}
O_B& & & & \\
 &\tilde{T}_{\mathrm{2}} & & & \mbox{\Large 0} \\
 & & O_A & &\\
 \mbox{\Large 0} & & &\tilde{T}_{\mathrm{1}} & \\
 & & & &O_B\\
\end{array}\right],\quad
\tilde{T}_{\alpha} = \left[ \begin{array}{cccc}
\bar{T}_{\alpha} & &  & \\
  & \bar{T}_{\alpha} &  &  \mbox{\Large 0} \\
 \mbox{\Large 0}  & & \ddots & \\
 &  &  & \bar{T}_{\alpha}\\
\end{array}\right],\quad
\bar{T}_{\alpha}=\left[ \begin{array}{cc}
-t_{\mathrm{f}} & 0\\
0 & - t_{\mathrm{f}} \\
\end{array}\right],\\
\check{H}_{\mathrm{f}} = \left[ \begin{array}{cc}
\hat{H}_{\mathrm{f}} & 0\\
0 & - \hat{H}^{\ast}_{\mathrm{f}} \\
\end{array}\right], \quad
\hat{H}_{\mathrm{f}} = \left[ \begin{array}{ccccc}
 O_B& & & & \\
 &\tilde{H}_2 & & & \mbox{\Large 0} \\
 & & O_A & &\\
 \mbox{\Large 0} & & &\tilde{H}_1 & \\
 & & & &O_B\\
\end{array}\right], \quad
\tilde{H}_{\alpha} = \left[ \begin{array}{ccccc}
\bar{H}_{\alpha} & \bar{T}_{\alpha}& &  & \mbox{\Large 0} \\
 \bar{T}_{\alpha} & \bar{H}_{\alpha} & \bar{T}_{\alpha} & & \\
  & \ddots& \ddots & \ddots & \\
  & & \bar{T}_{\alpha} & \bar{H}_{\alpha} & \bar{T}_{\alpha} \\
 \mbox{\Large 0} &  &  & \bar{T}_{\alpha} & \bar{H}_{\alpha}\\
\end{array}\right], \\
\bar{H}_{\alpha}=\left[ \begin{array}{cc}
4 t_{\mathrm{f}} - \mu_{\mathrm{f}} + M_{\alpha} \cos \varphi_\alpha & M_{\alpha} e^{- i \theta_\alpha} \sin \varphi_\alpha\\
M_{\alpha} e^{i \theta_\alpha} \sin \varphi_\alpha & 4 t_{\mathrm{f}} - \mu_{\mathrm{f}} - M_{\alpha} \cos \varphi_\alpha \\
\end{array}\right],
\end{gather}
and
\begin{gather}
\check{T}_{\mathrm{s}} = \left[ \begin{array}{cc}
\hat{T}_{\mathrm{s}} & \hat{T}_{\Delta} \\
\hat{T}_{\Delta} & - \hat{T}_{\mathrm{s}} \\
\end{array}\right], \quad
\hat{T}_\mathrm{s} = \left[ \begin{array}{cccc}
\bar{T}_\mathrm{s} & &  & \\
  & \bar{T}_\mathrm{s} &  &  \mbox{\Large 0} \\
 \mbox{\Large 0}  & & \ddots & \\
 &  &  & \bar{T}_\mathrm{s}\\
\end{array}\right],\quad
\hat{T}_{\Delta} = \left[ \begin{array}{cccc}
\bar{T}_{\Delta} & &  & \\
  & \bar{T}_{\Delta} &  &  \mbox{\Large 0} \\
 \mbox{\Large 0}  & & \ddots & \\
 &  &  & \bar{T}_{\Delta}\\
\end{array}\right],\\
\bar{T}_\mathrm{s}=\left[ \begin{array}{cc}
-t_{\mathrm{s}} & 0\\
0 & - t_{\mathrm{s}} \\
\end{array}\right],\quad
\bar{T}_{\Delta}=\left[ \begin{array}{cc}
0 & \frac{i \Delta_0}{2}\\
\frac{i \Delta_0}{2} & 0 \\
\end{array}\right],\\
\check{H}_{\mathrm{s}} = \left[ \begin{array}{cc}
\hat{H}_{\mathrm{s}} & \hat{\Delta} \\
-\hat{\Delta} & - \hat{H}_{\mathrm{s}} \\
\end{array}\right],\quad
\hat{H}_\mathrm{s} = \left[ \begin{array}{ccccc}
\bar{H}_\mathrm{s} & \bar{T}_\mathrm{s}& &  & \mbox{\Large 0} \\
 \bar{T}_\mathrm{s} & \bar{H}_\mathrm{s} & \bar{T}_\mathrm{s} & & \\
  & \ddots& \ddots & \ddots & \\
  & & \bar{T}_\mathrm{s} & \bar{H}_\mathrm{s} & \bar{T}_\mathrm{s} \\
 \mbox{\Large 0} &  &  & \bar{T}_\mathrm{s} & \bar{H}_\mathrm{s}\\
\end{array}\right], \quad
\hat{\Delta} = \left[ \begin{array}{ccccc}
0 & -\bar{\Delta}& &  & \mbox{\Large 0} \\
 \bar{\Delta} & 0 & -\bar{\Delta} & & \\
  & \ddots& \ddots & \ddots & \\
  & & \bar{\Delta} &0 & -\bar{\Delta} \\
 \mbox{\Large 0} &  &  & \bar{\Delta} & 0\\
\end{array}\right], \\
\bar{H}_\mathrm{s}=\left[ \begin{array}{cc}
4 t_{\mathrm{s}} - \mu_{\mathrm{s}} & 0\\
0 & 4 t_{\mathrm{s}} - \mu_{\mathrm{s}} \\
\end{array}\right],\quad
\bar{\Delta}=\left[ \begin{array}{cc}
0 & -\frac{\Delta_0}{2}\\
-\frac{\Delta_0}{2} & 0 \\
\end{array}\right].
\end{gather}
We represents $4N_\mathrm{s} \times 4N_\mathrm{s}$, $2N_\mathrm{s} \times 2N_\mathrm{s}$, $2N_\mathrm{f} \times 2N_\mathrm{f}$,
and $2 \times 2$ matrices by $\check{\cdots}$, $\hat{\cdots}$, $\tilde{\cdots}$ and $\bar{\cdots}$, respectively.
The $k \times k$ zero matrix is represented by $O_k$.
The $2N_\mathrm{f} \times 2N_\mathrm{f}$ matrix $\tilde{T}_{\alpha}$ ($\tilde{H}_{\alpha}$)
occupies from the $(2n_\alpha-1)$-th row to the $2N_\alpha$-th row of  $2N_\mathrm{s} \times 2N_\mathrm{s}$ matrix $\hat{T}_{\mathrm{f}}$ ($\hat{H}_{\mathrm{f}}$),
while other components of $\hat{T}_{\mathrm{f}}$ ($\hat{H}_{\mathrm{f}}$) are zero.

With this representation, the BdG equation describing the present junction is given by
\begin{gather}
 \check{T}(j-1) \boldsymbol{\psi}_{j-1}
+ \check{T}^{\dagger}(j) \boldsymbol{\psi}_{j+1}
+  \check{H} (j) \boldsymbol{\psi}_{j} = E \boldsymbol{\psi}_{j},\\
\boldsymbol{\psi}_{j} = [ \boldsymbol{u}_{j}, \boldsymbol{v}_{j} ]^{\mathrm{T}}, \quad
\boldsymbol{u}_{j}= [ \boldsymbol{u}_{j,1}, \boldsymbol{u}_{j,2}, \cdots, \boldsymbol{u}_{j,N_\mathrm{s}}]^{\mathrm{T}}, \quad
\boldsymbol{v}_{j}= [ \boldsymbol{v}_{j,1}, \boldsymbol{v}_{j,2}, \cdots, \boldsymbol{v}_{j,N_\mathrm{s}}]^{\mathrm{T}},\\
\boldsymbol{u}_{j,n} = [ \boldsymbol{u}_{\uparrow}(j,n), \boldsymbol{u}_{\downarrow}(j,n)]^{\mathrm{T}}, \quad
\boldsymbol{v}_{j,n} = [ \boldsymbol{v}_{\uparrow}(j,n), \boldsymbol{v}_{\downarrow}(j,n)]^{\mathrm{T}},
\end{gather}
with $-\infty \geq j \geq \infty$. 
For the FM lead $\alpha$ with $j \leq 0$, we can separate the BdG equation into the two Schr\"odinger equations as
\begin{gather}
\tilde{T}_{\alpha} \boldsymbol{u}^{\alpha}_{ j-1} + \tilde{T}_{\alpha}^{\dagger} \boldsymbol{u}^{\alpha}_{j+1}
+ \tilde{H}_{\alpha} \boldsymbol{u}^{\alpha}_{ j} = E \boldsymbol{u}^{\alpha}_{ j}, \label{eq:bdgeq_uf}\\
-\tilde{T}_{\alpha} \boldsymbol{v}^{\alpha}_{j-1} - \tilde{T}_{\alpha}^{\dagger} \boldsymbol{v}^{\alpha}_{ j+1}
 - \tilde{H}^{\ast}_{\alpha} \boldsymbol{v}^{\alpha}_{ j} = E \boldsymbol{v}^{\alpha}_{ j}, \label{eq:bdgeq_vf}\\
\boldsymbol{u}^{\alpha}_{ j}= [\boldsymbol{u}_{j,n_\alpha}, \boldsymbol{u}_{j,n_\alpha +1}, \cdots, \boldsymbol{u}_{j,N_\alpha}]^{\mathrm{T}}, \quad
\boldsymbol{v}^{\alpha}_{ j}= [\boldsymbol{v}_{j,n_\alpha}, \boldsymbol{v}_{j,n_\alpha +1}, \cdots, \boldsymbol{v}_{j,N_\alpha}]^{\mathrm{T}},
\end{gather}
where Eqs.~(\ref{eq:bdgeq_uf}) and (\ref{eq:bdgeq_vf}) respectively describe the electron and hole states in the FM lead $\alpha$.
For the SC segment with $j \geq2$, we obtain
\begin{align}
\check{T}_{s}  \boldsymbol{\psi}_{j-1} + \check{T}_{s}^{\dagger} \boldsymbol{\psi}_{j+1}
+ \check{H}_{s}  \boldsymbol{\psi}_{j} = E  \boldsymbol{\psi}_{j}. \label{eq:bdgeq_s}
\end{align}

By following the method shown in Ref.~[\onlinecite{ando_91}], we will derive an useful equation for solving the scattering problem.
As a preliminary step, we calculate the linearly independent solutions for Eqs.~(\ref{eq:bdgeq_uf}), (\ref{eq:bdgeq_vf}) and (\ref{eq:bdgeq_s}).
We first focus on the Eqs.~(\ref{eq:bdgeq_uf}) and (\ref{eq:bdgeq_vf}) describing the FM lead $\alpha$.
In the presence of translational symmetry in the $x$-direction, the solution of  Eqs.~(\ref{eq:bdgeq_uf}) and (\ref{eq:bdgeq_vf}) respectively satisfy
\begin{gather}
\boldsymbol{u}^{\alpha}_{ j} =
\lambda_\alpha \boldsymbol{u}^{\alpha}_{ j-1} =
\lambda_\alpha^j \boldsymbol{u}^{\alpha}_{0}, \label{eq:trns_uf} \\
\boldsymbol{v}^{\alpha}_{ j} =
\underline{\lambda}_\alpha \boldsymbol{v}^{\alpha}_{ j-1} =
\underline{\lambda}_\alpha^j \boldsymbol{v}^{\alpha}_{0}. \label{eq:trns_vf}
\end{gather}
By substituting Eq.~(\ref{eq:trns_uf}) into Eq.~(\ref{eq:bdgeq_uf}), and substituting Eq.~(\ref{eq:trns_vf}) into Eq.~(\ref{eq:bdgeq_vf}), we obtain
\begin{gather}
\lambda_\alpha \boldsymbol{u}^{\alpha}_{ j} = \left( \tilde{T}_{\alpha}^{\dagger}\right)^{-1} (E- \tilde{H}_{\alpha}) \boldsymbol{u}^{\alpha}_{ j}
- \left( \tilde{T}_{\alpha}^{\dagger}\right)^{-1} \tilde{T}_{\alpha} \boldsymbol{u}^{\alpha}_{ j-1}, \label{eq:bdgeq_uf2}\\
\underline{\lambda}_\alpha \boldsymbol{v}^{\alpha}_{ j} =
- \left( \tilde{T}_{\alpha}^{\dagger}\right)^{-1} (E+ \tilde{H}^{\ast}_{\alpha}) \boldsymbol{v}^{\alpha}_{ j}
- \left( \tilde{T}_{\alpha}^{\dagger}\right)^{-1} \tilde{T}_{\alpha} \boldsymbol{v}^{\alpha}_{ j-1}. \label{eq:bdgeq_vf2}
\end{gather}
By using Eqs.~(\ref{eq:trns_uf}) and (\ref{eq:bdgeq_uf2}), and using Eqs.~(\ref{eq:trns_vf}) and (\ref{eq:bdgeq_vf2}), we obtain the eigen equations
\begin{gather}
\left[\begin{array}{cc}
\left( \tilde{T}_{\alpha}^{\dagger}\right)^{-1} (E- \tilde{H}_{\alpha,\sigma}) &
-\left( \tilde{T}_{\alpha}^{\dagger}\right)^{-1} \tilde{T}_{\alpha}\\
1 & 0 \\
\end{array}\right]
\left[\begin{array}{c}
\boldsymbol{u}^{\alpha}_{ j}\\
\boldsymbol{u}^{\alpha}_{ j-1}\end{array} \right]
=\lambda_{\alpha}
\left[\begin{array}{c}
\boldsymbol{u}^{\alpha}_{ j}\\
\boldsymbol{u}^{\alpha}_{ j-1}\end{array} \right], \label{eq:andeq_uf}\\
\left[\begin{array}{cc}
-\left(\tilde{T}_{\alpha}^{\dagger}\right)^{-1} (E+\tilde{H}^{\ast}_{\alpha}) &
-\left( \tilde{T}_{\alpha}^{\dagger}\right)^{-1} \tilde{T}_{\alpha}\\
1 & 0 \\
\end{array}\right]
\left[\begin{array}{c}
\boldsymbol{v}^{\alpha}_{ j}\\
\boldsymbol{v}^{\alpha}_{ j-1}\end{array} \right]
=\underline{\lambda}_\alpha
\left[\begin{array}{c}
\boldsymbol{v}^{\alpha}_{ j}\\
\boldsymbol{v}^{\alpha}_{ j-1}\end{array} \right]. \label{eq:andeq_vf}
\end{gather}
By solving Eq.~(\ref{eq:andeq_uf}) numerically, we obtain the $4N_{\mathrm{f}}$ eigenstates, where
$2N_{\mathrm{f}}$ of them are right-going (left-going) eigenstates $\boldsymbol{\phi}_{\alpha, \eta}(+)$ [$\boldsymbol{\phi}_{\alpha, \eta}(-)$]
belonging with eigenvalue $\lambda_{\alpha, \eta}(+)$ [$\lambda_{\alpha, \eta}(-)$] for $\eta = 1$-$2N_{\mathrm{f}}$.
The right-going (left-going) propagating channel is characterized with $|\lambda_{\alpha, \eta}(\pm)|=1$ and $v_{\alpha, \eta}(+)>0$ ($v_{\alpha, \eta}(-)<0$),
where $v_{\alpha, \eta}(\pm)$ represents the group velocity of the electron state given by
\begin{align}
v_{\alpha, \eta}(\pm) = \mathrm{Im}
\left[ \boldsymbol{\phi}_{\alpha, \eta}^{\dagger}(\pm)
\left\{ \lambda_{\alpha, \eta}^{\ast}(\pm) \tilde{T}_{\alpha} - \lambda_{\alpha, \eta}(\pm) \tilde{T}_{\alpha}^{\dagger} \right\}
\boldsymbol{\phi}_{\alpha, \eta} (\pm) \right].
\end{align}
The right-going (left-going) evanescent channel is characterized with $|\lambda_{\alpha, \eta}(+)|<1$ ($|\lambda_{\alpha, \eta}(-)|>1$).
By using the eigenstates and eigenvelues of Eq.~(\ref{eq:andeq_uf}), we define two $2N_\mathrm{f} \times 2N_\mathrm{f}$ matrices
\begin{align}
&\tilde{U}_{\alpha} (\pm) = \left[ \boldsymbol{\phi}_{\alpha, 1}(\pm),
\boldsymbol{\phi}_{\alpha, 2}(\pm), \cdots,
\boldsymbol{\phi}_{\alpha, 2N_{\mathrm{f}}}(\pm) \right], \\
&\tilde{\Lambda}_{\alpha} (\pm) =
\mathrm{diag}\left[\lambda_{\alpha, 1}(\pm),
\lambda_{\alpha, 2}(\pm), \cdots,
\lambda_{\alpha, 2N_{\mathrm{f}}, \sigma}(\pm) \right].
\end{align}
As in the similar manner, we define two $2N_\mathrm{f} \times 2N_\mathrm{f}$ matrices by using the eigenstates and eigenvelues of Eq.~(\ref{eq:andeq_vf})
\begin{align}
&\tilde{\underline{U}}_{\alpha} (\pm) = \left[ \boldsymbol{\underline{\phi}}_{\alpha, 1}(\pm),
\boldsymbol{\underline{\phi}}_{\alpha, 2}(\pm), \cdots,
\boldsymbol{\underline{\phi}}_{\alpha, 2N_{\mathrm{f}}, \bar{\sigma}}(\pm) \right], \\
&\tilde{\underline{\Lambda}}_{\alpha} (\pm) =
\mathrm{diag}\left[\underline{\lambda}_{\alpha, 1}(\pm),
\underline{\lambda}_{\alpha, 2}(\pm), \cdots,
\underline{\lambda}_{\alpha, 2N_{\mathrm{f}}}(\pm) \right],
\end{align}
where $\boldsymbol{\underline{\phi}}_{\alpha, \eta}(+)$ [$\boldsymbol{\underline{\phi}}_{\alpha, \eta}(-)$]
represents the right-going (left-going) eigenstates corresponding to the eigenvalue $\underline{\lambda}_{\alpha, \eta}(+)$ [$\underline{\lambda}_{\alpha, \eta}(-)$]
for $\eta = 1$-$2N_{\mathrm{f}}$.
The right-going (left-going) propagating channel satisfies $|\underline{\lambda}_{\alpha, \eta}(\pm)|=1$
and $\underline{v}_{\alpha, \eta}(+)>0$ ($\underline{v}_{\alpha, \eta}(-)<0$) with
$\underline{v}_{\alpha, \eta}(\pm)$ representing the group velocity of the hole state given by
\begin{align}
\underline{v}_{\alpha, \eta}(\pm) = -\mathrm{Im}
\left[ \boldsymbol{\underline{\phi}}_{\alpha, \eta}^{\dagger}(\pm)
\left\{ \underline{\lambda}_{\alpha, \eta}^{\ast}(\pm) \tilde{T}_{\alpha} -
\underline{\lambda}_{\alpha, \eta}(\pm) \tilde{T}_{\alpha}^{\dagger} \right\}
\boldsymbol{\underline{\phi}}_{\alpha, \eta} (\pm) \right],
\end{align}
while the right-going (left-going) evanescent channel satisfies $|\underline{\lambda}_{\alpha, \eta}(+)|<1$ ($|\underline{\lambda}_{\alpha, \eta}(-)|>1$).
Any left-going and right-going electron (hole) states can be described by
the linear combination of $\boldsymbol{\phi}_{\alpha, \eta}(\pm)$ [$\boldsymbol{\underline{\phi}}_{\alpha, \eta}(\pm)$].
We here denote the left- and right-going electron states at $j=0$ with
\begin{gather}
\boldsymbol{u}^{\alpha}_{0}(\pm) = \sum_{\eta=1}^{2N_{\mathrm{f}}} c_{\alpha, \eta}(\pm) \boldsymbol{\phi}_{\alpha, \eta}(\pm)
= \tilde{U}_{\alpha} (\pm)  \boldsymbol{c}_{\alpha}(\pm), \label{eq:wave_uf0}\\
\boldsymbol{c}_{\alpha}(\pm)
= \left[ c_{\alpha, 1} (\pm), c_{\alpha, 2} (\pm), \cdots, c_{\alpha, 2N_{\mathrm{f}}} (\pm) \right]^{\mathrm{T}}, 
\end{gather}
and denote the left- and right-going hole states at $j=0$ with
\begin{gather}
\boldsymbol{v}^{\alpha}_{0}(\pm) =
\sum_{\eta=1}^{2N_{\mathrm{f}}} \underline{c}_{\alpha, \eta}(\pm) \boldsymbol{\underline{\phi}}_{\alpha, \eta}(\pm)
= \tilde{\underline{U}}_{\alpha} (\pm) \boldsymbol{\underline{c}}_{\alpha}(\pm), \label{eq:wave_vf0}\\
\boldsymbol{\underline{c}}_{\alpha}(\pm)
= \left[ \underline{c}_{\alpha, 1} (\pm), \underline{c}_{\alpha, 2} (\pm), \cdots, \underline{c}_{\alpha,2N_{\mathrm{f}}} (\pm) \right]^{\mathrm{T}},
\end{gather}
where  $c_{\alpha, \eta} (\pm)$ and $\underline{c}_{\alpha, \eta} (\pm)$ are expanding coefficients.
For $j \leq 0$, we can describe the left- and right-going states by
\begin{gather}
\boldsymbol{u}^{\alpha}_{ j}(\pm)
= \sum_{\eta=1}^{2N_{\mathrm{f}}} c_{\alpha,\eta}(\pm) \left\{  \lambda_{\alpha, \eta}(\pm) \right\}^{j} \boldsymbol{\phi}_{\alpha, \eta}(\pm)
= \tilde{U}_{\alpha} (\pm) \tilde{\Lambda}_{\alpha}^{j} (\pm) \boldsymbol{c}_{\alpha}(\pm) 
= \tilde{P}_{\alpha}^{j} (\pm)\boldsymbol{u}^{\alpha}_{0}(\pm), \label{eq:trns_uf(2)}\\
\tilde{P}_{\alpha}(\pm) = \tilde{U}_{\alpha} (\pm) \tilde{\Lambda}_{\alpha} \tilde{U}_{\alpha}^{-1} (\pm),\\
\boldsymbol{v}^{\alpha}_{ j}(\pm)
= \sum_{\eta=1}^{2N_{\mathrm{f}}} \underline{c}_{\alpha, \eta}(\pm)
\left\{  \underline{\lambda}_{\alpha, \eta}(\pm) \right\}^{j}
\boldsymbol{\underline{\phi}}_{\alpha, \eta}(\pm)
= \tilde{\underline{U}}_{\alpha} (\pm) \tilde{\underline{\Lambda}}_{\alpha}^{j} (\pm) \boldsymbol{c}_{\alpha}(\pm)
=  \tilde{\underline{P}}_{\alpha}^{j} (\pm) \boldsymbol{v}^{\alpha}_{0}(\pm), \label{eq:trns_vf(2)} \\
\tilde{\underline{P}}_{\alpha}(\pm) = \tilde{\underline{U}}_{\alpha} (\pm) \tilde{\underline{\Lambda}}_{\alpha} \tilde{\underline{U}}_{\alpha}^{-1} (\pm).
\end{gather}
Next, we calculate the the linearly independent solutions for Eq.~(\ref{eq:bdgeq_s}) describing the SC segment.
As similar to the analysis for the FM leads, we introduce the eigen equation
\begin{align}
\left[\begin{array}{cc}
\left( \check{T}_{s}^{\dagger}\right)^{-1} (E- \check{H}_{s}) &
-\left( \check{T}_{s}^{\dagger}\right)^{-1} \check{T}_{s}\\
1 & 0 \\
\end{array}\right]
\left[\begin{array}{c}
\boldsymbol{\psi}_{j}\\
\boldsymbol{\psi}_{j-1}\end{array} \right]
=\lambda_\mathrm{s}
\left[\begin{array}{c}
\boldsymbol{\psi}_{j}\\
\boldsymbol{\psi}_{j-1}\end{array} \right]. \label{eq:andeq_s}
\end{align}
By using the eigenstates and eigenvalues of of Eq.~(\ref{eq:andeq_s}), we construct the $4N_{\mathrm{s}} \times 4N_{\mathrm{s}}$ matrices
\begin{align}
&\check{U}_\mathrm{s} (\pm) = \left[ \boldsymbol{\phi}_{\mathrm{s},1}(\pm),
\boldsymbol{\phi}_{\mathrm{s}, 2}(\pm), \cdots,
\boldsymbol{\phi}_{\mathrm{s}, 4N_{\mathrm{s}}}(\pm) \right], \\
&\check{\Lambda}_\mathrm{s} (\pm) =
\mathrm{diag}\left[\lambda_{\mathrm{s},1}(\pm),
\lambda_{\mathrm{s},2}(\pm), \cdots,
\lambda_{\mathrm{s},4N_{\mathrm{s}}}(\pm) \right],
\end{align}
where $\boldsymbol{\phi}_{\mathrm{s}, \eta}(+)$ [$\boldsymbol{\phi}_{\mathrm{s}, \eta}(-)$]
represents the right-going (left-going) eigenstates belonging with eigenvelue $\lambda_{\mathrm{s}, \eta}(+)$ [$\lambda_{\mathrm{s}, \eta}(-)$] for $\eta=1$-$4N_{\mathrm{s}}$.
The right-going (left-going) states are characterized by $|\lambda_{\mathrm{s}, \eta}(+)|<0$ ($|\lambda_{\mathrm{s}, \eta}(-)|>0$)
or $v_{\mathrm{s}, \eta}(+)>0$ ($v_{\mathrm{s}, \eta}(-)<0$) with $|\lambda_{\mathrm{s}, \eta}(\pm)|=1$,
where $v_{\mathrm{s}, \eta}(\pm)$ represents the group velocity as
\begin{align}
v_{\mathrm{s}, \eta}(\pm) = \mathrm{Im}
\left[ \boldsymbol{\phi}_{\mathrm{s}, \eta}^{\dagger}(\pm)
\left\{ \lambda_{\mathrm{s}, \eta}^{\ast}(\pm) \check{T}_{s} - \lambda_{\mathrm{s}, \eta}(\pm) \check{T}_{s}^{\dagger} \right\}
\boldsymbol{\phi}_{\mathrm{s}, \eta} (\pm) \right].
\end{align}
We here represent the left- and right-going wave functions at $j=2$ as
\begin{gather}
\boldsymbol{\psi}_{2}(\pm) = \sum_{\eta=1}^{4N_{\mathrm{s}}}
c_{\mathrm{s}, \eta}(\pm) \boldsymbol{\phi}_{\mathrm{s}, \eta}(\pm)
= \check{U}_{\mathrm{s}} (\pm)  \boldsymbol{c}_{\mathrm{s}}(\pm), \label{eq:wave_s2}\\
\boldsymbol{c}_{\mathrm{s}}(\pm)
= \left[ c_{\mathrm{s}, 1} (\pm), c_{\mathrm{s}, 2} (\pm), \cdots, c_{\mathrm{s}, 4N_{\mathrm{s}}} (\pm) \right]^{\mathrm{T}}, 
\end{gather}
where $c_{\mathrm{s}, \eta} (\pm)$ is the expanding coefficient.
For $j \geq 1$, the wave function is given by
\begin{gather}
\boldsymbol{\psi}_{j}(\pm)
= \sum_{\eta=1}^{4N_{\mathrm{s}}} c_{\mathrm{s}, \eta}(\pm) \lambda^{j-2}_{\mathrm{s}, \eta}(\pm)  \boldsymbol{\phi}_{\mathrm{s}, \eta}(\pm)
= \check{U}_{\mathrm{s}} (\pm) \check{\Lambda}_{\mathrm{s}}^{j-2} (\pm) \boldsymbol{c}_{\mathrm{s}}(\pm), 
=\check{P}_{\mathrm{s}}^{j-2} (\pm)\boldsymbol{\psi}_{2}(\pm), \label{eq:trns_s}\\
\check{P}_{\mathrm{s}}(\pm) = \check{U}_{\mathrm{s}} (\pm) \check{\Lambda}_{\mathrm{s}} \check{U}_{\mathrm{s}}^{-1} (\pm).
\end{gather}

Let us now consider the scattering problem that electron states incident from the FM leads to the SC.
The electron wave function in the FM lead $\alpha$ at $j=0$ is represented as $\boldsymbol{u}^{\alpha}_0 = \boldsymbol{u}^{\alpha}_0(+) + \boldsymbol{u}^{\alpha}_0(-)$.
For $j=-1$, we obtain
\begin{align}
\boldsymbol{u}^{\alpha}_{ -1} &= \boldsymbol{u}^{\alpha}_{ -1}(+) + \boldsymbol{u}^{\alpha}_{ -1}(-) \nonumber\\
&=\tilde{P}_{\alpha}^{-1}(+) \boldsymbol{u}^{\alpha}_{ 0}(+) + \tilde{P}_{\alpha}^{-1}(-) \boldsymbol{u}^{\alpha}_{ 0}(-)
=\tilde{P}_{\alpha}^{-1}(+) \boldsymbol{u}^{\alpha}_{ 0}(+)
+ \tilde{P}_{\alpha}^{-1}(-) \left[ \boldsymbol{u}^{\alpha}_{ 0} - \boldsymbol{u}^{\alpha}_{ 0}(+) \right] \nonumber\\
&= \tilde{P}_{\alpha}^{-1}(-) \boldsymbol{u}^{\alpha}_{ 0}
+ \left[ \tilde{P}_{\alpha}^{-1}(+) - \tilde{P}_{\alpha}^{-1}(-) \right] \boldsymbol{u}^{\alpha}_{ 0}(+). \label{eq:wave_uf-1}
\end{align}
By substituting Eq.~(\ref{eq:wave_uf-1}) into Eq.~(\ref{eq:bdgeq_uf}), we deform the Schr\"odinger equation for the electron states at $j=0$ as
\begin{gather}
\tilde{T}_{\alpha}^{\dagger} \boldsymbol{u}^{\alpha}_{1}
+ \tilde{H}^{\mathrm{L}}_{\alpha} \boldsymbol{u}^{\alpha}_{0}
+ \tilde{Q}_{\alpha} \boldsymbol{u}^{\alpha}_{0} (+)
= E \boldsymbol{u}^{\alpha}_{0,\sigma}, \label{eq:scheq_ul},\\
\tilde{H}^{\mathrm{L}}_{\alpha} = \tilde{H}_{\alpha} + \tilde{T}_{\alpha} \tilde{P}_{\alpha}^{-1}(-), \quad
\tilde{Q}_{\alpha} = \tilde{T}_{\alpha} \left[ \tilde{P}_{\alpha}^{-1}(+) - \tilde{P}_{\alpha}^{-1}(-) \right].
\end{gather}
For $j \leq 0$, the wave function for the hole states consists of only left-going waves as $\boldsymbol{v}^{\alpha}_{j}=\boldsymbol{v}^{\alpha}_{j}(-)$.
Thus the hole wave function at $j=-1$ is deformed as
\begin{align}
\boldsymbol{v}^{\alpha}_{-1}=\boldsymbol{v}^{\alpha}_{-1}(-)
 = \tilde{\underline{P}}_{\alpha}^{-1} \boldsymbol{v}^{\alpha}_{0}(-)
=\tilde{\underline{P}}_{\alpha}^{-1} \boldsymbol{v}^{\alpha}_{0}. \label{eq:wave_vf-1}
\end{align}
By substituting Eq.~(\ref{eq:wave_vf-1}) into Eq.~(\ref{eq:bdgeq_vf}), we deform the Schr\"odinger equation for the hole states at $j=0$ as
\begin{gather}
-\tilde{T}_{\alpha}^{\dagger} \boldsymbol{v}^{\alpha}_{1} - \underline{\tilde{H}}^{\mathrm{L}}_{\alpha} \boldsymbol{v}^{\alpha}_{0}
= E \boldsymbol{v}^{\alpha}_{0}, \quad
\underline{\tilde{H}}^{\mathrm{L}}_{\alpha} = \tilde{H}^{\ast}_{\alpha}
+ \tilde{T}_{\alpha} \tilde{\underline{P}}_{\alpha}^{-1}(-) \label{eq:scheq_vl}.
\end{gather}
For the SC segment with $j \geq 1$, the wave function consists of only the right-going waves as $\boldsymbol{\psi}_{j}=\boldsymbol{\psi}_{j}(+)$.
Thus the wave function at $j =3$ is written as
\begin{align}
\boldsymbol{\psi}_{3} = \boldsymbol{\psi}_{3}(+)
= \check{P}_{\mathrm{s}} (+) \boldsymbol{\psi}_{2}(+)
= \check{P}_{\mathrm{s}} (+) \boldsymbol{\psi}_{2}.\label{eq:wave_s3}
\end{align}
By substituting Eq.~(\ref{eq:wave_s3}) into Eq.~(\ref{eq:bdgeq_s}), the BdG equation at $j=2$ is deformed as
\begin{align}
\check{T}_{\mathrm{s}}^{\dagger} \boldsymbol{\psi}_{1}
+ \check{H}^{\mathrm{R}} \boldsymbol{\psi}_{2}
= E \boldsymbol{\psi}_{2},\quad
\check{H}^{\mathrm{R}} = \check{H}_{\mathrm{s}} + \check{T}_{\mathrm{s}}^{\dagger} \check{P}_{\mathrm{s}}(+). \label{eq:bdgeq_s2}
\end{align}
By using Eqs.~(\ref{eq:scheq_ul}) , (\ref{eq:scheq_vl}) and (\ref{eq:bdgeq_s2}), we obtain a motion of equation for $0 \leq j \leq 2$
\begin{gather}
\left( E - \bar{H} \right)
\left[\begin{array}{c}
\boldsymbol{\psi}_{0}\\
\boldsymbol{\psi}_{1}\\
\boldsymbol{\psi}_{2} \end{array} \right]
=
\left[\begin{array}{c}
\check{Q} \boldsymbol{\psi}_{0}(+)\\
0\\
0 \end{array} \right], \quad
\bar{H} = \left[ \begin{array}{ccc}
\check{H}^{\mathrm{L}} & \check{T}_{\mathrm{f}}^{\dagger}& 0 \\
\check{T}_\mathrm{f} &\check{H}_{\mathrm{s}} & \check{T}_{\mathrm{s}}^{\dagger} \\
0 & \check{T}_{\mathrm{s}} & \check{H}^{\mathrm{R}}\\
\end{array}\right],\\
\boldsymbol{\psi}_{0} (+) =
\left[ \boldsymbol{u}_{0}(+), \boldsymbol{0}_{2N_{\mathrm{s}}} \right]^{\mathrm{T}}, \quad
\boldsymbol{u}_{0}(+) = \left[ \boldsymbol{0}_B ,
\boldsymbol{u}^2_{0}(+),
\boldsymbol{0}_A,
\boldsymbol{u}^1_{0}(+),
\boldsymbol{0}_B \right]^{\mathrm{T}},\\
\check{H}^{\mathrm{L}} = \left[ \begin{array}{cc}
\hat{H}^{\mathrm{L}} & 0\\
0 & - \underline{\hat{H}}^{\mathrm{L}} \\
\end{array}\right], \quad
\check{Q} = \left[ \begin{array}{cc}
\hat{Q} & 0\\
0 & 0 \\
\end{array}\right],\\
\hat{H}^{\mathrm{L}} = \left[ \begin{array}{ccccc}
 O_B& & & & \\
 &\tilde{H}^{\mathrm{L}}_2 & & & \mbox{\Large 0} \\
 & & O_A & &\\
 \mbox{\Large 0} & & &\tilde{H}^{\mathrm{L}}_1 & \\
 & & & &O_B\\
\end{array}\right], \quad
\underline{\hat{H}}^{\mathrm{L}} = \left[ \begin{array}{ccccc}
 O_B & & & & \\
 &\underline{\tilde{H}}^{\mathrm{L}}_2 & & & \mbox{\Large 0} \\
 & & O_A & &\\
 \mbox{\Large 0} & & &\underline{\tilde{H}}^{\mathrm{L}}_1 & \\
 & & & &O_B\\
\end{array}\right],\quad
\hat{Q} = \left[ \begin{array}{ccccc}
 O_B& & & & \\
 &\tilde{Q}_2 & & & \mbox{\Large 0} \\
 & & O_A & &\\
 \mbox{\Large 0} & & &\tilde{Q}_1 & \\
 & & & &O_B\\
\end{array}\right],
\end{gather}
where $\boldsymbol{0}_k$ is the zero vector with $k$ lines.
On the basis of this equation, we define the Green function obeying
\begin{gather}
\left[\begin{array}{c}
\boldsymbol{\psi}_{0}\\
\boldsymbol{\psi}_{1}\\
\boldsymbol{\psi}_{2} \end{array} \right]
= \bar{G}
\left[\begin{array}{c}
\check{Q}\boldsymbol{\psi}_{0}(+)\\
0\\
0 \end{array} \right], \label{eq:green}\\
\bar{G} = \left[ \begin{array}{ccc}
\check{G}(0,0) & \check{G}(0,1) & \check{G}(0,2) \\
\check{G}(1,0) & \check{G}(1,1) & \check{G}(1,2) \\
\check{G}(2,0) & \check{G}(2,1) & \check{G}(2,2)\\
\end{array}\right], \quad
\check{G}(j,j^{\prime}) = \left[ \begin{array}{cc}
\hat{G}(j,j^{\prime}) & \hat{F}(j,j^{\prime}) \\
\hat{\underline{F}}(j,j^{\prime}) & \hat{\underline{G}}(j,j^{\prime})
\end{array}\right]. 
\end{gather}
To calculate the reflection and transmission coefficients, we only need $\check{G}(0,0)$ and $\check{G}(1,0)$.
These matrix components can be easily calculated by using the recursive Green function technique as~\cite{fisher_81}
\begin{gather}
\check{G}(0,0)  = \check{G}^{\mathrm{L}} + \check{G}(0,1)  \check{T}_{\mathrm{f}} \check{G}^{\mathrm{L}}, \quad
\check{G}(1,0) = \check{G}(1,1) \check{T}_{\mathrm{f}}\check{G}^{\mathrm{L}},\quad
\end{gather}
where
\begin{gather}
\check{G}(1,1) = \left[ \check{G}_{\mathrm{s}}
- \check{T}_{\mathrm{s}}^{\dagger} \check{G}^{\mathrm{R}} \check{T}_{\mathrm{s}}
- \check{T}_{\mathrm{f}} \check{G}^{\mathrm{L}} \check{T}_{\mathrm{f}}^{\dagger} \right]^{-1},\quad
\check{G}(0,1) = \check{G}^{\mathrm{L}} \check{T}_{\mathrm{f}}^{\dagger}\check{G}(1,1),\\
\check{G}_{\mathrm{s}} = \left[ E - \check{H}_{\mathrm{s}} \right]^{-1}, \quad
\check{G}^{\mathrm{R}} = \left[ E - \check{H}^{\mathrm{R}} \right]^{-1},\quad
\check{G}_{\sigma}^{\mathrm{L}} = \left[ \begin{array}{cc}
\hat{G}^{\mathrm{L}} & 0 \\
 0 & \hat{\underline{G}}^{\mathrm{L}}
\end{array}\right],\\
\hat{G}^{\mathrm{L}} = \left[ \begin{array}{ccccc}
 O_B& & & & \\
 &\left[ E - \tilde{H}^{\mathrm{L}}_{2} \right]^{-1} & & & \mbox{\Large 0} \\
 & & O_A & &\\
 \mbox{\Large 0} & & &\left[ E - \tilde{H}^{\mathrm{L}}_{1} \right]^{-1}& \\
 & & & &O_B\\
\end{array}\right], \quad
\underline{\hat{G}}^{\mathrm{L}} = \left[ \begin{array}{ccccc}
 O_B& & & & \\
 &\left[ E + \underline{\tilde{H}}^{\mathrm{L}}_{2} \right]^{-1} & & & \mbox{\Large 0} \\
 & & O_A & &\\
 \mbox{\Large 0} & & &\left[ E + \underline{\tilde{H}}^{\mathrm{L}}_{1} \right]^{-1}& \\
 & & & &O_B\\
\end{array}\right].
\end{gather}

We here calculate the reflection coefficients.
By using Eqs.~(\ref{eq:wave_uf0}) and (\ref{eq:wave_vf0}), the wave function at $j=0$ is written by
\begin{gather}
\boldsymbol{\psi}_{0}=\left[\begin{array}{c}
\boldsymbol{u}_{0}(+)\\
0 \end{array} \right]
+
\left[\begin{array}{c}
\boldsymbol{u}_{0}(-)\\
\boldsymbol{v}_{0}(-) \end{array} \right]
=
\left[ \begin{array}{cc}
\hat{U}_{\mathrm{f}}(+) & 0\\
0 & 0 \\
\end{array}\right]
\left[\begin{array}{c}
\boldsymbol{c}_{\mathrm{f}}(+)\\
0 \end{array} \right]
+
\left[ \begin{array}{cc}
\hat{U}_{\mathrm{f}}(-) & 0\\
0 & \hat{\underline{U}}_{\mathrm{f}}(-) \\
\end{array}\right]
\left[\begin{array}{c}
\boldsymbol{c}_{\mathrm{f}}(-)\\
\underline{\boldsymbol{c}}_{\mathrm{f}}(-) \end{array} \right], \label{eq:wave_f0}\\
\boldsymbol{c}_{\mathrm{f}}(\pm) =
\left[ \boldsymbol{0}_B, \boldsymbol{c}_{2}(\pm),
\boldsymbol{0}_A, \boldsymbol{c}_{1}(\pm), \boldsymbol{0}_B \right]^{\mathrm{T}},\quad
\underline{\boldsymbol{c}}_{\mathrm{f}}(\pm) =
\left[ \boldsymbol{0}_B, \underline{\boldsymbol{c}}_{2}(\pm),
\boldsymbol{0}_A, \underline{\boldsymbol{c}}_{1}(\pm), \boldsymbol{0}_B \right]^{\mathrm{T}},\\
\hat{U}_{\mathrm{f}}(\pm) = \left[ \begin{array}{ccccc}
 O_B& & & & \\
 &\tilde{U}_{2}(\pm) & & & \mbox{\Large 0} \\
 & & O_A & &\\
 \mbox{\Large 0} & & &\tilde{U}_{1} (\pm) & \\
 & & & &O_B\\
\end{array}\right], \quad
\hat{\underline{U}}_{\mathrm{f}}(\pm) = \left[ \begin{array}{ccccc}
 O_B & & & & \\
 &\tilde{\underline{U}}_{2}(\pm) & & & \mbox{\Large 0} \\
 & & O_A & &\\
 \mbox{\Large 0} & & & \tilde{\underline{U}}_{1}(\pm) & \\
 & & & & O_B\\
\end{array}\right].
\end{gather}
From Eq.~(\ref{eq:green}), we also obtain 
\begin{align}
\boldsymbol{\psi}_{0} = \check{G}(0,0) \check{Q} \boldsymbol{\psi}_0 (+) 
=\check{G}(0,0) \left[ \begin{array}{cc}
\hat{Q}\hat{U}_{\mathrm{f}} (+) & 0 \\
 0 & 0
\end{array}\right]
\left[\begin{array}{c}
\boldsymbol{c}_{\mathrm{f}}(+)\\
0 \end{array} \right]. \label{eq:wave_f0(2)}
\end{align}
By combining Eqs.~(\ref{eq:wave_f0}) and (\ref{eq:wave_f0(2)}), we obtain
\begin{align}
\left[ \begin{array}{cc}
\hat{U}_{\mathrm{f}} (-) & 0 \\
 0 & \hat{\underline{U}}_{\mathrm{f}} (-) \\
\end{array}\right]
\left[\begin{array}{c}
\boldsymbol{c}_{\mathrm{f}}(-)\\
\boldsymbol{\underline{c}}_{\mathrm{f}}(-) \end{array} \right]
=
\left\{ \check{G}(0,0) \left[ \begin{array}{cc}
\hat{Q}\hat{U}_{\mathrm{f}} (+) & 0 \\
 0 & 0
\end{array}\right]
-\left[ \begin{array}{cc}
\hat{U}_{\mathrm{f}} (+) & 0 \\
 0 & 0
\end{array}\right] \right\}
\left[\begin{array}{c}
\boldsymbol{c}_{\mathrm{f}}(+)\\
0 \end{array} \right].
\end{align}
By using the matrices
\begin{align}
\hat{U}_{\mathrm{f}}^{\mathrm{Inv}} (-)=\left[ \begin{array}{ccccc}
 O_B& & & & \\
 &\tilde{U}_{2}^{-1}(\pm) & & & \mbox{\Large 0} \\
 & & O_A & &\\
 \mbox{\Large 0} & & &\tilde{U}_{1}^{-1}(\pm)& \\
 & & & &O_B\\
\end{array}\right], \quad
\hat{\underline{U}}_{\mathrm{f}}^{\mathrm{Inv}} (-)=\left[ \begin{array}{ccccc}
 O_B & & & & \\
 &\tilde{\underline{U}}_{2}^{-1}(\pm) & & & \mbox{\Large 0} \\
 & & O_A & &\\
 \mbox{\Large 0} & & &\tilde{\underline{U}}_{1}^{-1}(\pm)& \\
 & & & & O_B\\
\end{array}\right],
\end{align}
we obtain
\begin{gather}
\left[\begin{array}{c}
\boldsymbol{c}_{\mathrm{f}}(-)\\
\boldsymbol{\underline{c}}_{\mathrm{f}}(-) \end{array} \right]=
\left[ \begin{array}{cc}
\hat{{\cal R}}^{ee} & 0 \\
\hat{{\cal R}}^{he} & 0 \\
\end{array}\right]
\left[\begin{array}{c}
\boldsymbol{c}_{\mathrm{f}}(+)\\
0 \end{array} \right],\label{eq:rmat}\\
\hat{{\cal R}}^{ee} =
\hat{U}_{\mathrm{f}}^{\mathrm{Inv}} (-) \left\{ \hat{G}(0,0) \hat{Q} \hat{U}_{\mathrm{f}} (+) - \hat{U}_{\mathrm{f}} (+) \right\}
=\left[ \begin{array}{ccccc}
 O_B & & & & O_B \\
 &\tilde{{\cal R}}^{ee}_{22} & & \tilde{{\cal R}}^{ee}_{21} &  \\
 & & O_A & &\\
 &\tilde{{\cal R}}^{ee}_{12}& &\tilde{{\cal R}}^{ee}_{11}& \\
O_B & & & & O_B\\
\end{array}\right],\\
\hat{{\cal R}}^{he} = 
\hat{\underline{U}}_{\mathrm{f}}^{\mathrm{Inv}} (-) \hat{\underline{F}}(0,0) \hat{Q}\hat{U}_{\mathrm{f}} (+)
=\left[ \begin{array}{ccccc}
 O_B & & & & O_B \\
 &\tilde{{\cal R}}^{he}_{22} & & \tilde{{\cal R}}^{he}_{21} &  \\
 & & O_A & &\\
 &\tilde{{\cal R}}^{he}_{12}& &\tilde{{\cal R}}^{he}_{11} & \\
O_B & & & & O_B\\
\end{array}\right],
\end{gather}
where $\tilde{{\cal R}}^{ee}_{\beta\alpha}$ ($\tilde{{\cal R}}^{he}_{\beta\alpha}$) occupies
from $(2n_{\beta}-1)$-th to $2N_{\beta}$-th line and from $(2n_{\alpha}-1)$-th to $2N_{\alpha}$-th row of $\hat{{\cal R}}^{ee}$ ($\hat{{\cal R}}^{he}$).
From Eq.~(\ref{eq:rmat}), we find
\begin{gather}
c_{\beta,\zeta}(-) = \bar{r}^{ee}_{\beta\alpha} (\zeta; \eta) \; c_{\alpha,\eta}(+), \quad
\bar{r}^{ee}_{\beta\alpha} (\zeta; \eta)  = \tilde{{\cal R}}^{ee}_{\beta \alpha} |_{\zeta, \eta}\\
\underline{c}_{\beta,\zeta}(-)=
\bar{r}^{he}_{\beta\alpha} (\zeta; \eta) \; c_{\alpha,\eta}(+), \quad
\bar{r}^{he}_{\beta\alpha} (\zeta; \eta)  = \tilde{{\cal R}}^{he}_{\beta \alpha} |_{\zeta, \eta}.
\end{gather}
The normal (Andreev) reflection coefficients for the incident electron in the lead $\alpha$ belonging with the channel $\eta$
and the reflected electron (hole) in the lead $\beta$ belonging with the channel $\zeta$ is given by
\begin{gather}
r^{ee}_{\beta \alpha} (\zeta; \eta)=\left\{ \begin{array}{cl} 
\sqrt{v_{\beta,\zeta}(-) / v_{\alpha, \eta}(+)} \; \bar{r}^{ee}_{\beta\alpha} (\zeta; \eta) & \qquad \text{for} \quad v_{\alpha, \eta}(+) , v_{\beta,\zeta}(-) \neq 0 \\
0 & \qquad \text{otherwise}
\end{array}\right., \\
r^{he}_{\beta \alpha} (\zeta; \eta)=\left\{ \begin{array}{cl} 
\sqrt{\underline{v}_{\beta,\zeta}(-) / v_{\alpha, \eta}(+)} \; \bar{r}^{he}_{\beta\alpha} (\zeta; \eta)
& \qquad \text{for} \quad v_{\alpha, \eta}(+) , \underline{v}_{\beta,\zeta}(-) \neq 0 \\
0 & \qquad \text{otherwise}
\end{array}\right..
\end{gather}
We next calculate the transmission coefficients.
From Eq.~(\ref{eq:trns_s}), the wave function at $j=1$ is rewritten as
\begin{gather}
\boldsymbol{\psi}_{1} = \boldsymbol{\psi}_{1}(+) = \check{P}_{\mathrm{s}}^{-1} (+) \boldsymbol{\psi}_{2}(+)
= \check{U}_{\mathrm{s}} (+) \boldsymbol{d}_{\mathrm{s}}(+), \label{eq:wave_s1} \\
\boldsymbol{d}_{\mathrm{s}}(+) = \check{\Lambda}_{\mathrm{s}}^{-1} (+) \boldsymbol{c}_{\mathrm{s}}(+)
= \left[ d_{\mathrm{s}, 1} (\pm), d_{\mathrm{s}, 2} (\pm), \cdots, d_{\mathrm{s}, 4N_{\mathrm{s}}} (\pm) \right]^{\mathrm{T}}.
\end{gather}
From Eq.~ (\ref{eq:green}), we also find
\begin{align}
\boldsymbol{\psi}_{1} = \check{G}(1,0) \check{Q} \boldsymbol{\psi}_{0}(+) 
=\check{G}(1,0) \left[ \begin{array}{cc}
\hat{Q}\hat{U}_{\mathrm{f}} (+) & 0 \\
 0 & 0
\end{array}\right]
\left[\begin{array}{c}
\boldsymbol{c}_{\mathrm{f}}(+)\\
0 \end{array} \right]. \label{eq:wave_s1(2)}
\end{align}
By using Eqs.~(\ref{eq:wave_s1}) and (\ref{eq:wave_s1(2)}), we obtain
\begin{gather}
\boldsymbol{d}_{\mathrm{s}}(+)
=\check{{\cal T}}\left[\begin{array}{c}
\boldsymbol{c}_{\mathrm{f}}(+)\\ 0 \end{array} \right], \label{eq:tmat}\\
\check{{\cal T}} = \check{U}^{-1}_{\mathrm{s}} (+) \check{G}(1,0) \left[ \begin{array}{cc}
\hat{Q}\hat{U}_{\mathrm{f}} (+) & 0 \\
 0 & 0
\end{array}\right]
= \left[ \acute{O}_B, \breve{{\cal T}}_{2},\acute{O}_A, \breve{{\cal T}}_{1}, \acute{O}_{B+4N_\mathrm{s}} \right],
\end{gather}
where $\acute{O}_{k}$ represents $4N_{\mathrm{s}} \times k$ zero matrix.
The $4N_{\mathrm{s}} \times 2N_{\mathrm{f}}$ matrix $\breve{{\cal T}}_{\alpha}$
occupies from the $(2n_\alpha-1)$-th row to $2N_\alpha$-th row of the $4N_{\mathrm{s}} \times 4N_{\mathrm{s}}$ matrix $\check{{\cal T}}$.
From Eq.~(\ref{eq:tmat}), we obtain the relation
\begin{align}
 d_{\mathrm{s}, \zeta} (+) = \bar{t}_{\alpha} (\zeta; \eta) \; c_{\alpha, \eta} (+), \quad
\bar{t}_{\alpha} (\zeta; \eta) = \breve{{\cal T}}_{\alpha} |_{\zeta, \eta}.
\end{align}
The transmission coefficient for the incident electron in the FM lead $\alpha$ belonging with the channel $\eta$
and the out-going Bogoliubov quasiparticle in the SC belonging with the channel $\zeta$ is given by
\begin{align}
t_{\alpha} (\zeta; \eta)=\left\{ \begin{array}{cl} 
\sqrt{v_{\mathrm{s},\zeta}(+) / v_{\alpha, \eta}(+)} \; \bar{t}_{\alpha} (\zeta; \eta) & \qquad \text{for} \quad v_{\alpha, \eta}(+) , v_{\mathrm{s},\zeta}(-) \neq 0 \\
0 & \qquad \text{otherwise}
\end{array}\right..
\end{align}
These reflection and transmission coefficients satisfy the conservation low of
\begin{align}
\sum_{\beta=1,2} \;\sum_{\zeta=1}^{2N_\mathrm{f}}
\left\{ |r^{ee}_{\beta\alpha} (\zeta; \eta)|^2 + |r^{he}_{\beta\alpha}(\zeta; \eta)|^2 \right\} 
+ \sum_{\zeta=1}^{4N_\mathrm{s}} |t_{\alpha}(\zeta; \eta) |^2 =1.
\end{align}
To calculate the nonlocal conductance in the main text, 
we use the elastic co-tunneling and crossed Andreev reflection coefficients respectively given by
$r^{ee}_{\beta\alpha} (\zeta; \eta)$ and  $r^{he}_{\beta\alpha} (\zeta; \eta)$ with $\alpha \neq \beta$.

%
\subsection{Wave functions}
%
In this section, we explain the calculation method for the spacial profile of wave functions shown in Fig.~4 in the main text.
From Eqs.~(\ref{eq:trns_uf(2)}), (\ref{eq:trns_vf(2)}) and (\ref{eq:wave_f0}), the wave function in the FM segment $j\leq0$ is described as
\begin{gather}
\boldsymbol{\psi}_{j}=\left[\begin{array}{c}
\boldsymbol{u}_{j}(+)\\
0 \end{array} \right]
+
\left[\begin{array}{c}
\boldsymbol{u}_{j}(-)\\
\boldsymbol{v}_{j}(-) \end{array} \right]
=
\left[ \begin{array}{cc}
\hat{U}_{\mathrm{f}}(+) \hat{\Lambda}_{\mathrm{f}}^j(+) & 0\\
0 & 0 \\
\end{array}\right]
\left[\begin{array}{c}
\boldsymbol{c}_{\mathrm{f}}(+)\\
0 \end{array} \right]
+
\left[ \begin{array}{cc}
\hat{U}_{\mathrm{f}}(-) \hat{\Lambda}_{\mathrm{f}}^j(-) & 0\\
0 & \hat{\underline{U}}_{\mathrm{f}}(-) \hat{\underline{\Lambda}}_{\mathrm{f}}^j(-) \\
\end{array}\right]
\left[\begin{array}{c}
\boldsymbol{c}_{\mathrm{f}}(-)\\
\underline{\boldsymbol{c}}_{\mathrm{f}}(-) \end{array} \right],\\
\hat{\Lambda}_{\mathrm{f}}(\pm)=\left[ \begin{array}{ccccc}
 O_B & & & & \\
 &\tilde{\Lambda}_{2}(\pm) & & & \mbox{\Large 0} \\
 & & O_A & &\\
 \mbox{\Large 0} & & &\tilde{\Lambda}_{1} (\pm)& \\
 & & & &O_B\\
\end{array}\right], \quad
\hat{\underline{\Lambda}}_{\mathrm{f}}(\pm)=\left[ \begin{array}{ccccc}
 O_B & & & & \\
 & \tilde{\underline{\Lambda}}_{2}(\pm) & & & \mbox{\Large 0} \\
 & & O_A & &\\
 \mbox{\Large 0} & & & \tilde{\underline{\Lambda}}_{1}(\pm) & \\
 & & & & O_B\\
\end{array}\right].
\end{gather}
By using Eq.~(\ref{eq:rmat}), we obtain
\begin{align}
\boldsymbol{\psi}_{j}=\left[ \begin{array}{cc}
\hat{U}_{\mathrm{f}}(+) \hat{\Lambda}_{\mathrm{f}}^j(+) & 0\\
0 & 0 \\
\end{array}\right]
\left[\begin{array}{c}
\boldsymbol{c}_{\mathrm{f}}(+)\\
0 \end{array} \right]
+
\left[ \begin{array}{cc}
\hat{U}_{\mathrm{f}}(-) \hat{\Lambda}_{\mathrm{f}}^j(-) \hat{{\cal R}}^{ee} & 0\\
\hat{\underline{U}}_{\mathrm{f}}(-) \hat{\underline{\Lambda}}_{\mathrm{f}}^j(-) \hat{{\cal R}}^{he} & 0\\
\end{array}\right]
\left[\begin{array}{c}
\boldsymbol{c}_{\mathrm{f}}(+)\\
0 \end{array} \right]. \label{eq:wave_fall}
\end{align}
From Eqs.~(\ref{eq:trns_s}) and (\ref{eq:wave_s1}), the wave function for the SC segment $j \geq 1$ is written by
\begin{align}
\boldsymbol{\psi}_{j} = \check{U}_{\mathrm{s}} (+)  \check{\Lambda}_{\mathrm{s}}^{(j-1)}(+)  \boldsymbol{d}_{\mathrm{s}}(+).
\end{align}
By using Eq~(\ref{eq:tmat}), we find
\begin{align}
\boldsymbol{\psi}_{j} = \check{U}_{\mathrm{s}} (+)  \check{\Lambda}_{\mathrm{s}}^{(j-1)}(+)  \check{{\cal T}}\left[\begin{array}{c}
\boldsymbol{c}_{\mathrm{f}}(+)\\ 0 \end{array} \right]. \label{eq:wave_sall}
\end{align}

Let us focus on the wave function $\boldsymbol{\psi}_{\eta_0, j}$ belonging with the incoming channel $\eta_0$ in the FM lead $\alpha$.
To calculate $\boldsymbol{\psi}_{\eta_0, j}$, we set the expanding coefficient $\boldsymbol{c}_{\mathrm{s}}(+)$ as
\begin{align}
c_{\alpha, \eta} (+) = \left\{ \begin{array}{cl} 
1 & \text{for} \quad \eta = \eta_0 \\
0 & \text{otherwise}
\end{array}\right., \qquad
c_{\beta, \eta} (+) = 0 \quad \text{for all $\eta$} \label{eq:const},
\end{align}
with $\alpha \neq \beta$.
By substituting Eqs.~(\ref{eq:const}) into Eq.~(\ref{eq:wave_fall}) and (\ref{eq:wave_sall}), we finally obtain
\begin{gather}
\boldsymbol{\psi}_{\eta_0, j} = \left\{ \begin{array}{cl} 
\boldsymbol{\psi}^{\alpha}_{\eta_0, j} & \text{for FM lead $\alpha$ \qquad ($j \leq 0$, $n_{\alpha} \leq n \leq N_{\alpha}$)} \\
\boldsymbol{\psi}^{\beta}_{\eta_0, j} &  \text{for FM lead $\beta$ \qquad ($j \leq 0$, $n_{\beta} \leq n \leq N_{\beta}$) }\\
\boldsymbol{\psi}^{\mathrm{s}}_{\eta_0, j} &  \text{for superconductor \qquad ($j \geq 1$, $n_{\mathrm{s}} \leq n \leq N_{\mathrm{s}}$)}\\
0 &  \text{otherwise}
\end{array}\right.,\\
\boldsymbol{\psi}^{\alpha}_{\eta_0, j}  = \left[\begin{array}{c}
\boldsymbol{u}^{\alpha}_{\eta_0, j} \\
\boldsymbol{v}^{\alpha}_{\eta_0, j} \end{array} \right]
= \left[ \begin{array}{c}
\left[ \tilde{U}_{\alpha}(+) \tilde{\Lambda}_{\alpha}^j(+) \right]_{\text{$\eta_0$-th row}} \\
0\\
\end{array}\right]
+
\left[ \begin{array}{c}
\left[\tilde{U}_{\alpha}(-) \tilde{\Lambda}_{\alpha}^j(-) \tilde{{\cal R}}^{ee}_{\alpha \alpha}\right]_{\text{$\eta_0$-th row}}\\
\left[\tilde{\underline{U}}_{\alpha}(-) \tilde{\underline{\Lambda}}_{\alpha}^j(-) \tilde{{\cal R}}^{he}_{\alpha \alpha} \right]_{\text{$\eta_0$-th row}}\\
\end{array}\right],\\
\boldsymbol{\psi}^{\beta}_{\eta_0, j}  = \left[\begin{array}{c}
\boldsymbol{u}^{\beta}_{\eta_0, j} \\
\boldsymbol{v}^{\beta}_{\eta_0, j} \end{array} \right]
= \left[ \begin{array}{c}
\left[\tilde{U}_{\beta}(-) \tilde{\Lambda}_{\beta}^j(-) \tilde{{\cal R}}^{ee}_{\beta \alpha} \right]_{\text{$\eta_0$-th row}}\\
\left[\tilde{\underline{U}}_{\beta}(-) \tilde{\Lambda}_{\beta}^j(-)\tilde{{\cal R}}^{he}_{\beta \alpha}\right]_{\text{$\eta_0$-th row}}\\
\end{array}\right],\\
\boldsymbol{\psi}^{\mathrm{s}}_{\eta_0, j}  = \left[\begin{array}{c}
\boldsymbol{u}^{\mathrm{s}}_{\eta_0, j} \\
\boldsymbol{v}^{\mathrm{s}}_{\eta_0, j} \end{array} \right]=
\left[\check{U}_{\mathrm{s}} (+) \check{\Lambda}_{\mathrm{s}}^{(j-1)}(+) \breve{{\cal T}}_{\alpha} \right]_{\text{$\eta_0$-th row}},
\end{gather}
where
\begin{gather}
\boldsymbol{u}^{\alpha}_{\eta_0, j}
= \left[ \boldsymbol{u}_{\eta_0, j,n_{\alpha}},  \boldsymbol{u}_{\eta_0, j, n_{\alpha}+1},
\cdots, \boldsymbol{u}_{\eta_0, j, N_{\alpha}} \right]^{\mathrm{T}}, \quad
\boldsymbol{v}^{{\alpha}}_{\eta_0, j}
= \left[ \boldsymbol{v}_{\eta_0, j,n_{\alpha}},  \boldsymbol{v}_{\eta_0, j, n_{\alpha}+1},
\cdots, \boldsymbol{v}_{\eta_0, j, N_{\alpha}} \right]^{\mathrm{T}},\\
\boldsymbol{u}^{\beta}_{\eta_0, j}
= \left[ \boldsymbol{u}_{\eta_0, j,n_{\beta}},  \boldsymbol{u}_{\eta_0, j, n_{\beta}+1},
\cdots, \boldsymbol{u}_{\eta_0, j, N_{\beta}} \right]^{\mathrm{T}}, \quad
\boldsymbol{v}^{{\beta}}_{\eta_0, j}
= \left[ \boldsymbol{v}_{\eta_0, j,n_{\beta}},  \boldsymbol{v}_{\eta_0, j, n_{\beta}+1},
\cdots, \boldsymbol{v}_{\eta_0, j, N_{\beta}} \right]^{\mathrm{T}},\\
\boldsymbol{u}^{\mathrm{s}}_{\eta_0, j}
= \left[ \boldsymbol{u}_{\eta_0, j,1},  \boldsymbol{u}_{\eta_0, j, 2},
\cdots, \boldsymbol{u}_{\eta_0, j, N_\mathrm{s}} \right]^{\mathrm{T}}, \quad
\boldsymbol{v}^{\mathrm{s}}_{\eta_0, j}
= \left[ \boldsymbol{v}_{\eta_0, j,1},  \boldsymbol{v}_{\eta_0, j, 2},
\cdots, \boldsymbol{v}_{\eta_0, j, N_\mathrm{s}} \right]^{\mathrm{T}},\\
\boldsymbol{u}_{\eta_0, j,n}
\left[ u_{\eta_0,\uparrow}(j,n),  u_{\eta_0,\downarrow}(j,n) \right]^{\mathrm{T}}, \quad
\boldsymbol{v}_{\eta_0, j,n}
\left[ v_{\eta_0,\uparrow}(j,n),  v_{\eta_0,\downarrow}(j,n) \right]^{\mathrm{T}}.
\end{gather}
In the main text, we show $\psi_{\eta_{\mathrm{M}}}(\boldsymbol{r})
= [u_{\eta_{\mathrm{M}},\uparrow}(\boldsymbol{r}), u_{\eta_{\mathrm{M}},\downarrow}(\boldsymbol{r}),
v_{\eta_{\mathrm{M}},\uparrow}(\boldsymbol{r}), v_{\eta_{\mathrm{M}},\downarrow}(\boldsymbol{r})]^{\mathrm{T}}$ at zero energy
belonging with the incoming channel $\eta_{\mathrm{M}}$ having the largest value of $\sum_{\zeta} | r^{he}_{21}(\zeta;\eta) |^{2}$ among all $\eta$.

\section{Transmission and Reflection Probabilities \\ at a Chiral $p$-wave Superconductor/Normal-metal Interface}

\begin{figure}[hhhh]
\begin{center}
\includegraphics[width=0.7\textwidth]{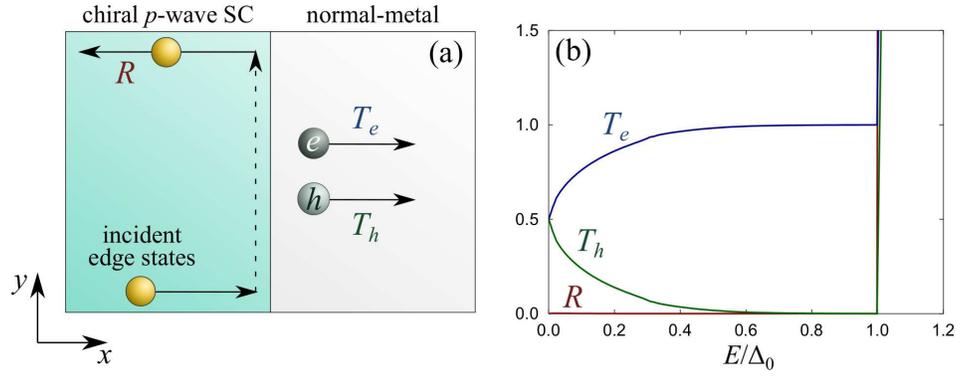}
\caption{(a)Schematic image of scattering processes at the chiral $p$-wave superconductor/normal-metal interface,
where the chiral edge states incident from the lower edge of the superconductor to the normal-metal.
(b)Reflection and transmission probabilities as a function of energy of the incident modes from the superconductor.}
\label{fig:supplement2}
\end{center}
\end{figure}

In the Blonder-Tinkham-Klapwijk formalism, we assume that the charge currents carried by the chiral Majorana edge states moving towards
the inside of the superconducting segment ($x = + \infty$) are absorbed into the ideal electrode attached to the superconductor.
In this section, to support this assumption, we calculate the reflection and transmission probabilities in a chiral $p$-wave superconductor/normal-metal (SN) junction
as shown in Fig.~\ref{fig:supplement2}(a).
We consider the present junction on the two-dimensional lattice model with the lattice constant $a_0$.
A lattice site is indicated by a vector $\boldsymbol{r}=j\boldsymbol{x}+m\boldsymbol{y}$, 
where $\boldsymbol{x}$ ($\boldsymbol{y}$) is the vector in the $x$ ($y$) direction with $|\boldsymbol{x}|= |\boldsymbol{y}|=a_0$.
The chiral $p$-wave superconductor (normal-metal) occupies $j \leq 0$ ($j \geq 1$) and $- M \leq m \leq M$.
In the $y$ direction, we apply the hard-wall boundary condition.
The present junction is described by the Bogoliubov-de Gennes Hamiltonian
\begin{align}
H = &-t \sum_{j=-\infty}^{\infty} \sum_{m=-M}^{M}
\left[c^{\dagger}_{\boldsymbol{r}+\boldsymbol{x}}c_{\boldsymbol{r}}+c^{\dagger}_{\boldsymbol{r}}c_{\boldsymbol{r}+\boldsymbol{x}} \right]
-t \sum_{j=-\infty}^{\infty} \sum_{m=-M}^{M-1}
\left[c^{\dagger}_{\boldsymbol{r}+\boldsymbol{y}}c_{\boldsymbol{r}}+c^{\dagger}_{\boldsymbol{r}}c_{\boldsymbol{r}+\boldsymbol{y}} \right] \nonumber\\
& + \frac{i \Delta_0}{4}\sum_{j=-\infty}^{-1} \sum_{m=-M}^{M}
\left[c^{\dagger}_{\boldsymbol{r}+\boldsymbol{x}}c^{\dagger}_{\boldsymbol{r}}- c^{\dagger}_{\boldsymbol{r}}c^{\dagger}_{\boldsymbol{r}+\boldsymbol{x}} \right]
+ \mathrm{H.c.} \nonumber\\
& - \frac{\chi \Delta_0}{4}\sum_{j=-\infty}^{0} \sum_{m=-M}^{M-1}
\left[c^{\dagger}_{\boldsymbol{r}+\boldsymbol{y}}c^{\dagger}_{\boldsymbol{r}}- c^{\dagger}_{\boldsymbol{r}}c^{\dagger}_{\boldsymbol{r}+\boldsymbol{y}} \right]
+ \mathrm{H.c.},
\end{align}
where $c_{\boldsymbol{r}}^{\dagger}$($c_{\boldsymbol{r}}$) represents
the creation (annihilation) operator of an electron at the site $\boldsymbol{r}$,
 $t$ denotes the nearest-neighbor hopping integral, and $\mu$ is the chemical potential.
The amplitude and chirality of the pair potential in the superconducting segment are represented by $\Delta_0$ and $\chi$ ($=1$ or $-1$), respectively.
In what follows, we fix several parameters as
$\mu=2.0t$, $\Delta = 0.1t$, $\chi = -1$, and $M=100$.
With $\chi = -1$, the chiral Majorana edge states of the chiral $p$-wave superconductor incident from the lower edge ($m=-M$) of the superconducting segment to the SN interface
as shown in Fig.~\ref{fig:supplement2}(a).
By using the lattice Green functions technique, we calculate the reflection and transmission probabilities defined as
\begin{align}
R(E) = \sum_{\zeta_s,\eta} |r_{\zeta_s,\eta}(E)|^2, \quad T_{e(h)}(E) = \sum_{\zeta_n,\eta} |t^{e(h)}_{\zeta_n,\eta}(E)|^2.
\end{align}
The reflection coefficient at energy $E$ is given by $r_{\zeta_s,\eta}(E)$,
where the index $\eta$ labels the incident channel from the superconducting segment and the index $\zeta_s$ labels the outgoing channel in the superconductor. 
The transmission coefficient from the quasi-particle states in the superconductor to the electron (hole) states in the normal segment is represented by $t^{e(h)}_{\zeta_n,\eta}(E)$,
where the index $\zeta_n$ labels the outgoing channel in the normal-metal segment.
With the energy below the superconducting gap (i.e., $E<\Delta$), there is only one incident channel corresponding to the chiral Majorana edge state at the lower edge.
Therefore,  the reflection probability $R$ with $E<\Delta$ corresponds to the scattering processes
that the incident chiral Majorana edge states are reflected to the superconducting segment as the backward chiral Majorana edge states at the upper edge
as shown Fig.~\ref{fig:supplement2}(a).
In Fig.~2(b), we show $R$, $T_e$ and $T_h$ as a function of energy of incident states from the superconducting segment.
With $E < \Delta_0$, we find the important relations of $R=0.0$ and $T_e+T_h=1.0$,
which imply that the incident chiral Majorana edge states are always scattered into the attached normal-metal.
Although the normal-metal does not describe the ideal electrode straightforwardly, this result strongly support the assumption of the BTK formalism that
the chiral Majorana edge states moving toward $x = + \infty$ are always absorbed into the ideal electrode and never circle around the edge of superconductor.
\clearpage

\end{document}